\documentclass[preprint2,amsmath,trackchanges,resetfootnote]{aastex702}
\usepackage{amsmath}
\usepackage{subcaption}
\usepackage{bm}
\usepackage{placeins}
\begin{document}

\title{The Influence of the Inclination of the Accretion Disk in  Black Hole X-Ray Binaries on the Energy Spectrum Characteristics}

\author{Aijun Dong}
\affiliation{School of Physics and Electronic Science, Guizhou Normal University, Guiyang 550001, People’s Republic of China}
\affiliation{Guizhou Provincial Key Laboratory of Radio Astronomy and Data Processing, Guizhou Normal University, Guiyang 550001, People’s Republic of China}
\email{aijdong@gznu.edu.cn}

\author{Shangyao Jing}
\affiliation{School of Physics and Electronic Science, Guizhou Normal University, Guiyang 550001, People’s Republic of China}
\email{jingsy@gznu.edu.cn}

\author{Weilong He}
\affiliation{School of Physics and Electronic Science, Guizhou Normal University, Guiyang 550001, People’s Republic of China}
\email{hewl@gznu.edu.cn}

\begin{abstract}

During the ascending and descending phases of some black hole X-ray binaries(BH XRBs), it was found that the X-ray Eddington ratio ($ \rm L_{X-ray(3-25keV)}/L_{Edd} $) and the photon spectral index $\Gamma$ had an anticorrelation when the X-ray Eddington ratio was less than a specific value, and a positive correlation when it was greater than that value. Moreover, the cross points of the anticorrelation and positive correlation are different for different BH XRBs ($ \rm L_{X-ray(3-25keV)}/L_{Edd} $ of cross point is between $ 0.1 \% $ and $ 10\% $). The reason why the cross points are different for different BH XRBs were still uncertain. So we processed the light curve and energy spectrum of 15 outbursts of eight BH XRBs during their descending phases. We found that for the same BH XRB, the values of the X-ray Eddington ratio at the cross point are basically consistent for different outburst times. However, for different BH XRBs, there is a positive linear correlation between the X-ray Eddington ratio at the cross point and the cosine of the accretion disk inclination. Therefore, We think that the geometric effect of the inclination has a significant impact on the evolution of the energy spectrum of the BH XRBs.

\end{abstract}



\section{Introduction} 

Black hole X-ray binaries(BH XRBs) are an important system for studying the properties related to BHs. This system is mainly composed of a stellar-mass BH with a mass ranging from a few to $ \rm \sim 20 M_\odot$ , and a massive companion star \citep{Fender2012}. The matter of the companion star fills the roche lobe and flows towards the BH through the first lagrange point. Due to the angular momentum carried by the matter flowing into the BH, it forms an accretion disk around the BH and releases high-energy X-ray radiation as a result of the release of gravitational energy of the matter \citep{Fender2012}. However, except for a few rare cases where the BH XRBs have stable and powerful X-ray emissions (such sources are called ``persistent sources"), the rest of the BH XRBs can only be observed to have X-ray emissions during their outbursts (such sources are called ``transient sources"). For these BH XRBs, most of the time, they are in a quiescent state, with a low luminosity ($ \rm L_X/L_{Edd} \lesssim 10^-5 $). The X-ray energy spectrum is dominated by power-law components (the photon spectral index $ \Gamma \sim 1.5 - 2.0 $). As matter accumulates on the accretion disk, they will undergo an outburst, and their X-ray luminosity will increase by several orders of magnitude during the outburst until it approaches the Eddington luminosity ($ \rm L_{Edd} = 10^{38} \times M_{BH}/M_{\odot} $, where $ \rm M_{BH} $ is the BH mass). During the entire outburst of the X-ray binary, in the initial stage of the outburst, the X-ray binary first enters the low-hard state (LHS), at which the X-ray energy spectrum is dominated by power-law components, with a weak disk component contribution, and the spectrum is relatively hard ($ \Gamma \sim 1.5 - 2.0 $), and the luminosity is low. The X-ray emission has been suggested to come from an advection-dominated accretion flow (ADAF; \citep{Abramowicz1995,Narayan1995,Liu2007,Liu2010,Yuan2014}). As the luminosity increases, the X-ray binary gradually enters the intermediate state (IMS). The X-ray energy spectrum is dominated by both power-law components and accretion disk components, and the IMS can be divided into the hard IMS and the soft IMS according to the proportion of high-energy X-ray photons produced. Immediately after that, the X-ray binary will enter the high-soft state (HSS) as the proportion of high-energy photons decreases. At this time, the X-ray energy spectrum is dominated by the accretion disk component, with a weak power-law component contribution, and the spectrum is relatively soft ($ \Gamma > 2.0 $), and the luminosity is very high. The disk and power-law components come from an optically thick, standard accretion thin disk ( Shakura-Sunyaev Disk; \citep{1973A&A....24..337S})  and an optically thin, hot corona surrounding it. While the canonical picture often assumes a corona located above and below the disk, the exact geometry of the corona whether it is a compact lamppost, an extended slab-like structure, or a more complex configuration remains an open and actively debated question \citep{Liu2022, Zhang2024, Fan2025}. After this, the luminosity further increases, and the X-ray binary will enter an extremely steep ($ \Gamma > 2.4 $) very high state (VHS) (or steep-power-law state, or intermediate state). Then, the X-ray binary will return to the low-hard state and eventually to the quiescent state (see \citep{Remillard2006,Done2007,Fender2012,2013FrPhy...8..630Z}, for reviews).

The photon spectral index (hereafter: $ \Gamma $) can reflect the properties of hot plasma (corona/ADAF/jet) without considering the contribution of possibly cold accretion disks. Therefore, it is also widely used in the study of the same physics in BHs of different sizes and in the physics of the same coronal matter. \citet{Wu2008} studied the energy spectral evolution during the descending phase of some X-ray binaries and pointed out that in the descending phase of the X-ray outbursts of BH XRBs, the X-ray Eddington ratio (hereafter: $ \xi_X = \rm L_{X-ray(3-25keV)}/L_{Edd} $) has an anticorrelation with $ \Gamma $ when it is less than a specific $ \xi_X $, and a positive correlation when it is greater than that $ \xi_X $. In other words, this specific $ \xi_X $ is the cross point of the anticorrelation and positive correlation of $\Gamma - \xi_X$ evolution diagram (hereafter: $ \rm \xi_{X_{cross \ point}}$). However, the positions of the cross points of different BH XRBs are different ($ \rm  \xi_{X_{cross \ point}}  = 0.1\% \sim 10\% $) \citep{Wu2008}. After this, \citet{Liu2019} also simultaneously studied the evolution of the $ \Gamma $ of the ascending and descending phases of the outbursts of some BH XRBs with respect to the total X-ray flux (hereafter: $ \rm F_{3-25keV} $ ). They also found that for both the ascending and descending phases of the outburst, the relationship diagrams between $ \rm F_{3-25keV} $ and $ \Gamma $ both exhibited anticorrelation and positive correlation, and the anticorrelation and positive correlation of the two had both similarities and differences. Specifically, the anticorrelations of the ascending phase and the descending phase had the same trend, but the positive correlations have different slopes \citep{Liu2019}. Currently, there is no relatively good explanation for why there are different trends on the positive correlation \citep{Liu2019}. And \citet{Liu2019} also discuss the reason why the positions of the cross points of different BH XRBs are different; they mentioned that inclination may have some effect.

The inclination of the accretion disk is an important parameter for BH XRBs and is widely used in calculating the mass function of the BH mass \citep{Orosz1998,Orosz2011,Torres2020,YanesRizo2022}. Different active galactic nuclei (AGNs) also have different observed inclinations. \citet{Xu2015} constructed a model of the accretion disk-corona for AGN; through the radiation transfer theory, it was calculated that for AGN, there is a positive relationship between the X-ray luminosity and the cosine value of the observed inclination of that AGN \citep{Xu2015}. Previous studies have found that BH XRBs and AGN have many similar structures and characteristics, in other words, BH XRBs can be seen as the ``small-scale AGN," which has a much shorter variation time scale \citep{2003MNRAS.345.1057M,Uttley2005,McHardy2006,McHardy2009}, and some works also pointed out that the X-ray spectrum of BH XRBs may be inclination-dependent \citep{MuozDarias2013,Heil2015}. In addition to the energy spectrum, multiple observational characteristics of BH XRBs have also been found to be sensitive to the inclination. In terms of time-varying characteristics, \citet{Vincentelli2026} quantified the significant linear correlation between the amplitude of Type-C quasi-periodic oscillations (QPOs) and the jet inclination for the first time through a systematic analysis of the amplitude evolution of QPOs in the Rossi X-ray Timing Explorer(RXTE) and Insight-HXMT archives. This result supports that QPOs originate from geometrically related modulation mechanisms (such as precession heat flow or the projection effect of the corona), rather than isotropic luminosity oscillations \citep{Ingram2019,Vincentelli2026}. This result is also consistent with theoretical predictions from precessing ring models \citep{Schnittman2006} and spiral instability models \citep{Varniere2016}, which both find that QPO modulation amplitude increases with inclination. In terms of polarization, the observations by IXPE make it possible to directly measure the polarization degree of BH XRBs, and the polarization degree itself carries key information about the inclination of the system. \citet{1980A&A....86..121S} pointed out theoretically long ago that the linear polarization degree generated by electron scattering in the disk-like Compton cloud follows the classical Chandrasekhar distribution and is highly sensitive to the inclination; recent IXPE analyses of sources such as 4U 1957+115 further confirm that the polarization degree can be used as an independent method for estimating the inclination, independent of the dynamical method \citep{Marra2024,Titarchuk2025}.

In this work, we obtain the cross point of the anticorrelation and positive correlation of $\Gamma - \xi_X $ evolution diagram by performing spectral fitting and further processing on the energy spectral data of 15 outbursts of 8 BH XRBs, and then explore the correlation between this and the inclination of the BH XRBs. The structure of this paper is as follows: The second part will introduce the data and data processing, the third part will introduce the results of curve fitting, the fourth part will be the discussion, and the fifth part will be the conclusion. 

\section{Data and Data processing} \label{s2}

\subsection{Data}
We refer to the low-mass X-ray binary star sample compiled by \citet{2021MNRAS.507.5507A}. In their work, they used all the available data from the RXTE and we selected the required samples from it. The selection criteria were as follows: {\uppercase\expandafter{\romannumeral1}}. The type of the outburst is one that has undergone a complete outburst type (i.e., completed the entire state evolution); {\uppercase\expandafter{\romannumeral2}}. The source has a definite inclination range (with an error of $ 10^{\circ} $); {\uppercase\expandafter{\romannumeral3}}. The source has a definite mass and distance; {\uppercase\expandafter{\romannumeral4}}. The data during the outburst period of the source is available for download. Based on these selection criteria, eight sources were selected as samples. According to the relevant papers reviewed, the mass, distance, and inclination of these sources were recorded. Their specific information is shown in Table \ref{t1}.

\begin{deluxetable}{lcccccc}
\tablewidth{0pt}
\tablecaption{Parameter information for each source \label{t1}}
\tablehead{
\colhead{Source} & \colhead{Outburst year} & \colhead{$ \rm N_H $} & \colhead{BH mass} & \colhead{Distance} & \colhead{Inclination} \\
\colhead{} & \colhead{} & \colhead{$ (\rm \times 10^{22}cm^{-2}) $} & \colhead{$ \rm(M_\odot) $} & \colhead{$ \rm(kpc) $} & \colhead{$ \rm(degree) $}
}
\startdata
4U 1543-47 & 2002 & $0.4^{[1]}$ & $5.1\pm 2.4^{[2]}$ & $7.5\pm 1.0^{[3]}$ & $20.7\pm 1.5^{[4]}$ \\
\hline
XTE J1752-223 & 2009 & $0.46^{[5]}$ & $9.5\pm 1.5^{[5]}$ & $6.0\pm 2.0^{[6]}$ & $35.0\pm 4.0^{[7]}$ \\
\hline
GX 339-4 & 2002 & $0.5^{[8]}$ & $7.5\pm 0.8^{[9]}$ & $10.0\pm 2.0^{[10]}$ & $50.0\pm 10.0^{[10]}$ \\
         & 2004 &             &                    &                      &                       \\
         & 2007 &             &                    &                      &                       \\
         & 2010 &             &                    &                      &                       \\
\hline
XTE J1859+226 & 1999 & $0.2^{[11]}$ & $7.8\pm 1.9^{[12]}$ & $4.2\pm 0.5^{[13]}$ & $50.0\pm 3.0^{[12]}$ \\
\hline
GRO J1655-40 & 2005 & $0.8^{[14]}$ & $6.0\pm 0.4^{[15]}$ & $3.2\pm 0.5^{[16]}$ & $70.2\pm 1.9^{[17]}$ \\
\hline
XTE J1550-564 & 1998 & $2.0^{[18]}$ & $9.1\pm 0.6^{[19]}$ & $4.4\pm 0.5^{[19]}$ & $74.7\pm 3.0^{[19]}$ \\
              & 2000 &              &                     &                     &                      \\
\hline
H 1743-322 & 2003 & $1.8^{[20]}$ & $11.2\pm 1.7^{[21]}$ & $8.5\pm 0.8^{[22]}$ & $75.0\pm 3.0^{[22]}$ \\
           & 2004 &              &                      &                     &                      \\
           & 2005 &              &                      &                     &                      \\
           & 2009 &              &                      &                     &                      \\
\hline
MAXI J1659-152 & 2011 & $0.3^{[23]}$ & $5.7\pm 1.8^{[24]}$ & $6.0\pm 2.0^{[24]}$ & $75.0\pm 5.0^{[24]}$
\enddata
\tablecomments{Refrence:[1]:\cite{Dong2020}; [2]:\cite{1998ApJ...499..375O}; [3]:\cite{2004MNRAS.354..355J}; [4]:\cite{2003IAUS..212..365O}; [5]:\cite{Shaposhnikov2010}; [6]:\cite{Ratti2012}; [7]:\cite{2018ApJ...864...25G}; [8]:\cite{2023AdSpR..71.3508D}; [9]:\cite{2011IAUS..275..327C}; [10]:\cite{2019MNRAS.488.1026Z}; [11]:\cite{Nandi2018}; [12]:\cite{2022MNRAS.517.1476Y}; [13]:\cite{2009ApJ...699..453S}; [14]:\cite{2016MNRAS.463..615K}; [15]:\cite{2003MNRAS.339.1031S}; [16]:\cite{1995Natur.375..464H}; [17]:\cite{2001ApJ...554.1290G}; [18]:\cite{2002ApJ...567.1057T}; [19]:\cite{2011ApJ...730...75O}; [20]:\cite{2011MNRAS.414..677C}; [21]:\cite{2017ApJ...834...88M}; [22]:\cite{2012ApJ...745L...7S}; [23]:\cite{2015ApJ...803...59D}; [24]:\cite{Torres2020}.}
\end{deluxetable}

\subsection{Data processing}
\subsubsection{Energy Spectrum Fitting}
\label{s2.2.1}

We utilized all the \emph{RXTE} observational data of the selected BH XRBs and carried out data processing for spectral fitting. We used the data reduction tools in {\small HEASoft} (Version 6.30)\footnote{\url{https://heasarc.gsfc.nasa.gov/docs/software/lheasoft/}}, following the data reduction description in the \emph{RXTE Cook Book}\footnote{\url{https://heasarc.gsfc.nasa.gov/docs/xte/recipes/cook_book.html}}. First, we created a new filter file using {\small XTEFILE}. Then, we created a GTI file using {\small MAKETIME} with the standard screening criteria recommended for faint sources: elevation above the earth's limb {\small ELV} $> 10^{\circ}$ and pointing offset {\small OFFSET} $< 0.02^{\circ}$ \citep{Liu2019}. To exclude data contaminated by the South Atlantic Anomaly (SAA) passage, we applied the constraint {\small TIME\_SINCE\_SAA} $> 30$~min (or {\small TIME\_SINCE\_SAA} $< 0$) \citep{Jahoda2006}. We also applied the electron ratio cut {\small ELECTRON2} $< 0.1$ to reject periods of elevated particle background \citep{Jahoda2006}. Based on the count rate greater than 40 in the PHA file, we selected the bright background model for addition. When the count rate was less than 40, we selected the dark background model for addition. Next, we used {\small GRPPHA} to extract the spectra with background and added $1\%$ systematic error to all spectra. The data reduction and spectral extraction were automated to handle each observation in the same way. To maintain consistency, only the data from the proportional counter array (PCA) PCU-2 was used, as this special unit remained open throughout the \emph{RXTE} operation and could be consistently utilized throughout the data archive process. Additionally, it is also the best calibration unit on \emph{RXTE}. The data energy range selection was within $\rm 3-25~keV$.

The spectral fitting was carried out using {\small XSPEC}(version 12.10.1)\footnote{\url{https://heasarc.gsfc.nasa.gov/docs/software/xspec/}}. Two models, $ \tt phabs*powerlow $ and $ \tt phabs*(diskbb + powerlow) $, were adopted to study the evolution of the thermal and nonthermal components during the descending stage of the outburst. These models allow for the study of multicolor blackbody components and power-law components. Since the $\rm 6 \ keV $ $\rm Fe $ $\alpha $ line has been detected in the spectra of many X-ray binaries \citep{1995A&AS..109....9G,Fabian2000}, Gaussian lines with a consistent energy value of $ \rm 6.4 \ keV $ were included in both of the above models. This resulted in the creation of a total of four models: $ \tt phabs*(gaussian + powerlow)$, $ \tt phabs*(diskbb + gaussian + powerlow) $, $ \tt phabs*powerlow $, and $ \tt phabs*(diskbb + powerlow) $. Additionally, the hydrogen column density $ \rm N_H $ for each source was fixed and shown in the Table \ref{t1}. The goodness of fitting was evaluated using the chi-square value $ (0.5 \le {\chi^2} \le 1.5) $, and the systematic error was set at $ 0.5\% $. Assuming that all sources are isotropic, the X-ray flux was fitted within the energy range of $ \rm 3-25 keV $. Based on the obtained X-ray flux, the X-ray luminosity $ \rm (L_x = 4{\bm\pi} \times D^2 \times Flux) $ was calculated, where $\rm D $ represents the distance from the observer to the source, in units of centimeters. During the fitting process, the power-law components of some data points in $\rm HSS$ were very weak, and the $ \Gamma $ of some data in $\rm VHS$ could not be  constrained. Therefore, the data points with a power-law fraction lower than $\rm 10\%$ and points without constrained $ \Gamma $ were ignored.

\subsubsection{The photon spectral index and the X-ray Eddington ratio evolution diagram fitting and linear relationship fitting}
\label{s2.2.2}
Before fitting the $\Gamma - \xi_X$ evolution graph, we need to conduct a certain screening of the data obtained from the energy spectrum fitting. To avoid the selection effect, we have set the same screening criteria for the data obtained from the energy spectral fitting of each outburst of each source. The specific screening criteria are as follows:

\begin{enumerate}
\item The data used for the fitting of our $\Gamma - \xi_X$ evolution diagram are the data of the light curve during the descending phase of each outburst of each source, since more observational data in the descending phase can enable us to obtain a more complete $\Gamma - \xi_X$ evolution diagram. During our data processing, the definition of the descending phase is that in the light curve of the outburst, those data points with MJD greater than or equal to the MJD corresponding to the maximum point of the total flux  in the X-ray band ($\rm F_{3-25KeV}$) of the light curve.
\item We filtered out some abnormal data points. The abnormal data points include: {\uppercase\expandafter{\romannumeral1}}. Data points with extremely large $\Gamma$ and their errors. If the spectral fitting is not good enough during the energy spectrum fitting, it may lead to an extremely large $\Gamma$ or error. During our data processing, we consider data points with a $\Gamma$ greater than 3.5 or an error greater than 0.5 as abnormal data points; {\uppercase\expandafter{\romannumeral2}}. If the $\Gamma$ of a point is greater than the maximum $\Gamma$ corresponding to the the leftmost and rightmost regions of $\Gamma - \xi_X$ evolution diagram, then this point will also be regarded as an abnormal data point. According to the results of \citet{Wu2008} and \citet{Liu2019}, for the $\Gamma - \xi_X$ evolution diagram, at low $\xi_X$, the $\Gamma$ is inversely proportional to the $\xi_X$, and at $\xi_X$, the $\Gamma$ is directly proportional to the $\xi_X$. Therefore, for the $\Gamma$ in the middle area of the $\Gamma - \xi_X$ evolution diagram, it is reasonable that the $\Gamma$ should be smaller than the maximum $\Gamma$ at both ends. Therefore, some data points with excessive $\Gamma$ in the middle area need to be filtered out as abnormal data points; {\uppercase\expandafter{\romannumeral3}}. Data points where the fraction of the power-law component is less than $ 10\% $, these data points are difficult to constrain their $\Gamma$ in the energy spectrum fitting due to the excessively small fraction of power-law components.
\item We need to filter out the data points located within the ``flare". Because during the ``flare", there may be physical mechanisms completely different from normal outbursts \citep{2022ApJ...933...55C}, which will have a certain impact on the fitting of the $\Gamma - \xi_X$ evolution diagram. The quantitative judgment criteria for the ``flare" are as follows: If there are four data points with continuous increases in the total flux in X-ray band ($\rm F_{3-25KeV}$), we will consider that there is a ``flare" here. The first data point among these four points is regarded as the starting point of the ``flare". Then, the data points where the $\rm F_{3-25KeV}$ and MJD are both greater than $\rm F_{3-25KeV}$ and MJD of the ``flare" starting point will be regarded as data points located within the ``flare" and will be filtered out.
\item Finally, we need to filter out the data points whose $\Gamma$ do not change significantly with the $\xi_X$ and are located on the right side of the evolution graph (i.e., in the early descending stage of the light curve). In \citet{Liu2019}, they found that before the $\xi_X$ and $\Gamma$ began to evolve in a positive correlation or negative correlation during the descending stage of the light curve, there was a period during which the $\Gamma$ hardly changed with $\xi_X$. This period would seriously interfere with our subsequent fitting of the positive correlation and negative correlation between $\xi_X$ and $\Gamma$. Therefore, the data points in this period need to be filtered out. To quantitatively filter out this part of the data points, we choose to use two different functions to fit the $\Gamma - \xi_X$ evolution diagram composed of the remaining data after the above three data filtering steps. These two functions are
\begin{align}
f_1(x) &=
\begin{cases}
k_1(x-x_0)+y_0 & x<x_0\\
k_2(x-x_0)+y_0 & x \ge x_0
\end{cases}\label{eq1}\\[6pt]
f_2(x) &=
\begin{cases}
k_1(x-x_0)+y_0 & x<x_0\\
k_2(x-x_0)+y_0 & x_0 \leq x < x_1\\
k_2(x_1-x_0)+y_0 & x \ge x_1
\end{cases}\label{eq2}
\end{align}
\noindent Among them, $f_1(x)$ is a simple linear piecewise function, $\rm k_1$ and $\rm k_2$ represent the slopes of two linear functions, while $\rm x_0$ and $\rm y_0$ are the x-coordinate and y-coordinate of the cross point the anticorrelation and positive correlation of the $\Gamma - \xi_X $ evolution diagram; $f_2(x)$ is a linear piecewise function combined with a constant function, the first two parts of $f_2(x)$ are same with $f_1(x)$, and the constant function is equal to $\rm k2(x1-x0) + y0$, the cross point of the linear piecewise function and the constant function is $ (x_1, \ k2(x1-x0) + y0)$ . We use these two functions to fit the same data, and then calculate the corresponding Akaike information criterion (AIC) of these two functions. The specific calculation formula of AIC is: 
\begin{equation}
\mathrm{AIC} = n\ln\left(\frac{\mathrm{SSE}}{n}\right) + 2p
\end{equation}
\noindent where $\rm p$ is the number of fitting parameters, $\rm n$ is the number of fitting data points, and  the ``SSE" is the squares of the residuals between the real data and the corresponding values of the best-fitted curve \citep{Akaike1974}. Generally, if the AIC of one model is 10 less than that of the other model, we consider that the difference in data fitting between the two models is significant and the fitting function with the larger AIC can be directly rejected \citep{burnham2002model}. For the two functions we used, if the AIC of $f_1(x)$ minus the AIC of $f_2(x)$ is greater than 10, we consider that there is a stage in the data where $\Gamma$ almost does not change with $\xi_X$. To retain more data as much data as possible, we will exclude the data that the $\xi_X$ is greater than the value of $\rm x_1+0.1$ and the $\rm x_1$ is one of the best-fitted parameter of $f_2(x)$. If the AIC of $f_1(x)$ minus the AIC of $f_2(x)$ is less than 10, we consider that the data does not have the stage where $\Gamma$ almost does not change with $\xi_X$.
\end{enumerate}

For the $\Gamma - \xi_X $ evolution diagram fitting, we use $f_1(x)$ (Equation \eqref{eq1}) which is already been introduced. After defining our objective function and the relevant fitting parameters, we can use the $\rm scipy.optimize.curve\_fit $ \footnote{\url{https://docs.scipy.org/doc/scipy/reference/generated/scipy.optimize.curve_fit.html}} module to perform the least squares fitting on our data. We use the Markov Chain Monte Carlo method (MCMC method) to obtain the error of the fitting parameters. The corresponding $\Gamma - \xi_X$ evolution diagram fitting results will be presented in Section \ref{s3.1}.

After fitting the $\Gamma - \xi_X$ evolution diagram, we can obtain the size of the X-axis  of the cross point ($\rm \xi_{X_{cross \ point}}$) in the $\Gamma - \xi_X$ evolution diagram for each source during each outburst. Next, we can use the linear function $\rm y = k*x + b $ to fit the existing data to determine whether there is a linear relationship between the cosine value of inclination and $\rm \xi_{X_{cross \ point}}$ and the specific size of the parameters of the linear relationship. The fitting process is basically similar to that of the $\Gamma - \xi_X$ evolution diagram. So, no further explanation is needed.

\section{Result}
\subsection{The photon spectral index and X-ray Eddington ratio evolution diagram fitting result}
\label{s3.1}

\begin{figure*}[htbp]
    \centering
    \subfloat[4U 1543-47]{\includegraphics[width=0.28\textwidth]{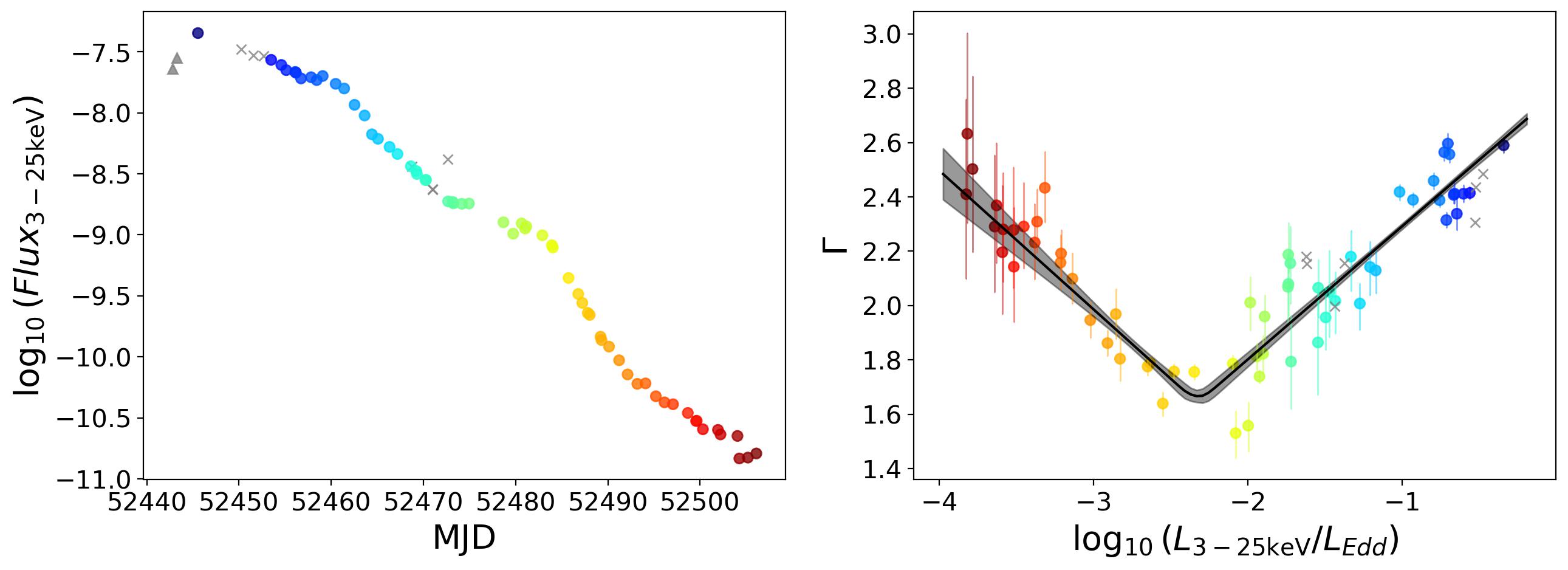}}
    \subfloat[XTE J1752-223]{\includegraphics[width=0.28\textwidth]{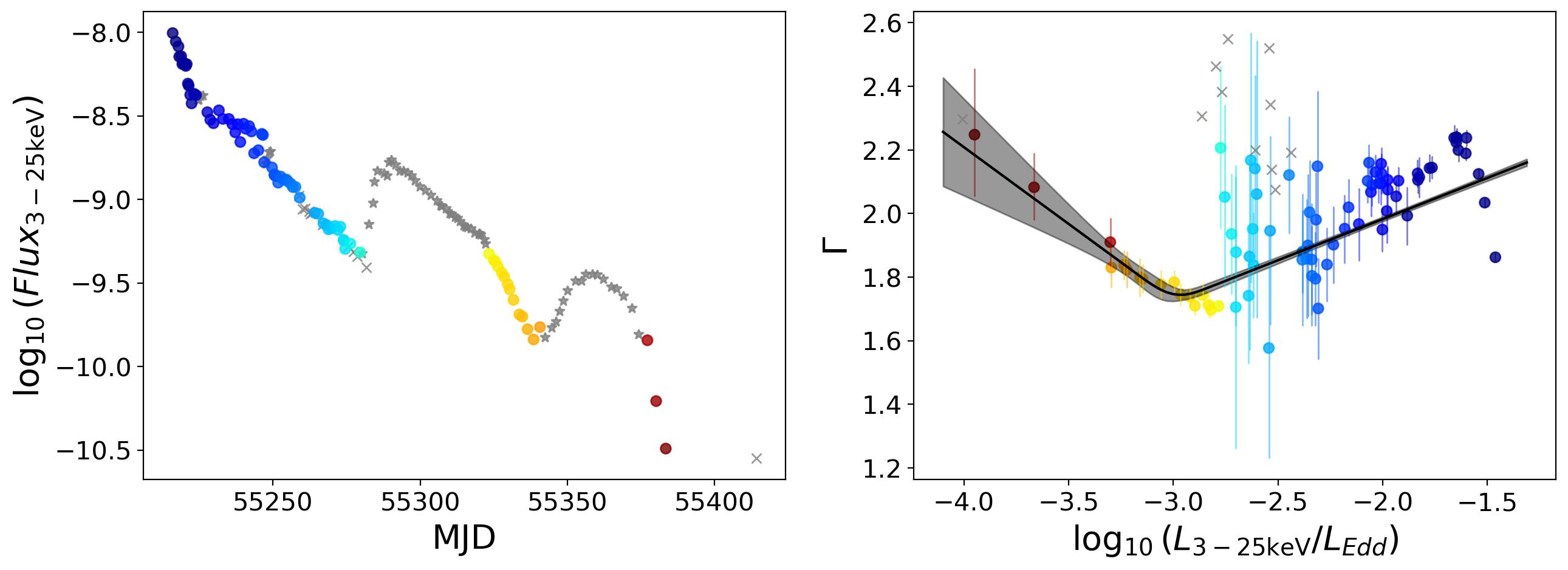}}
    \subfloat[GX 339-4 2002]{\includegraphics[width=0.28\textwidth]{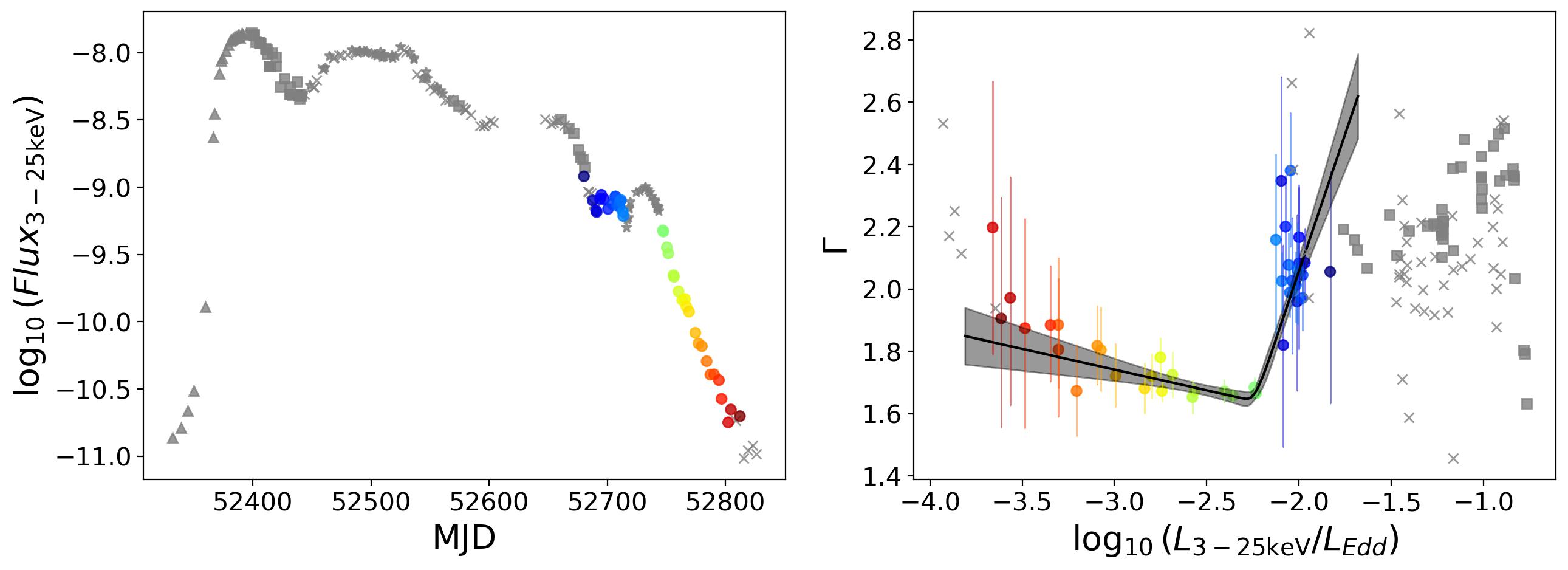}}
    \\
    \subfloat[GX 339-4 2004]{\includegraphics[width=0.28\textwidth]{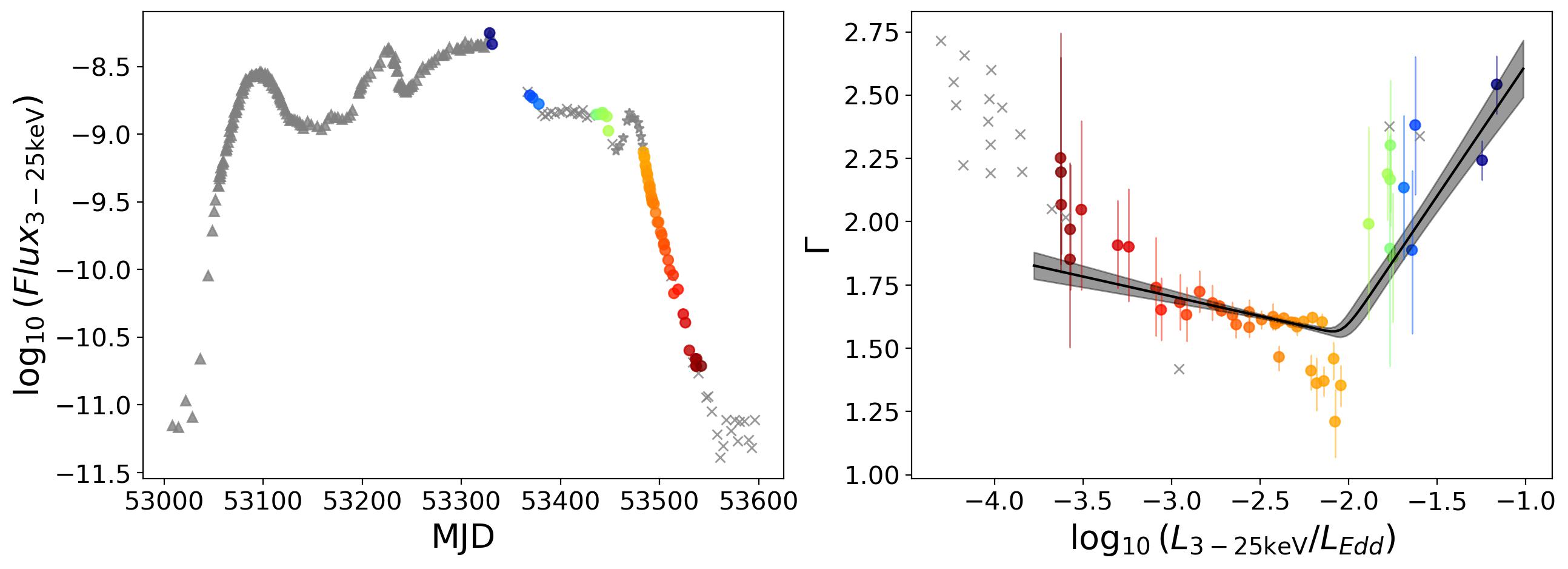}}
    \subfloat[GX 339-4 2007]{\includegraphics[width=0.28\textwidth]{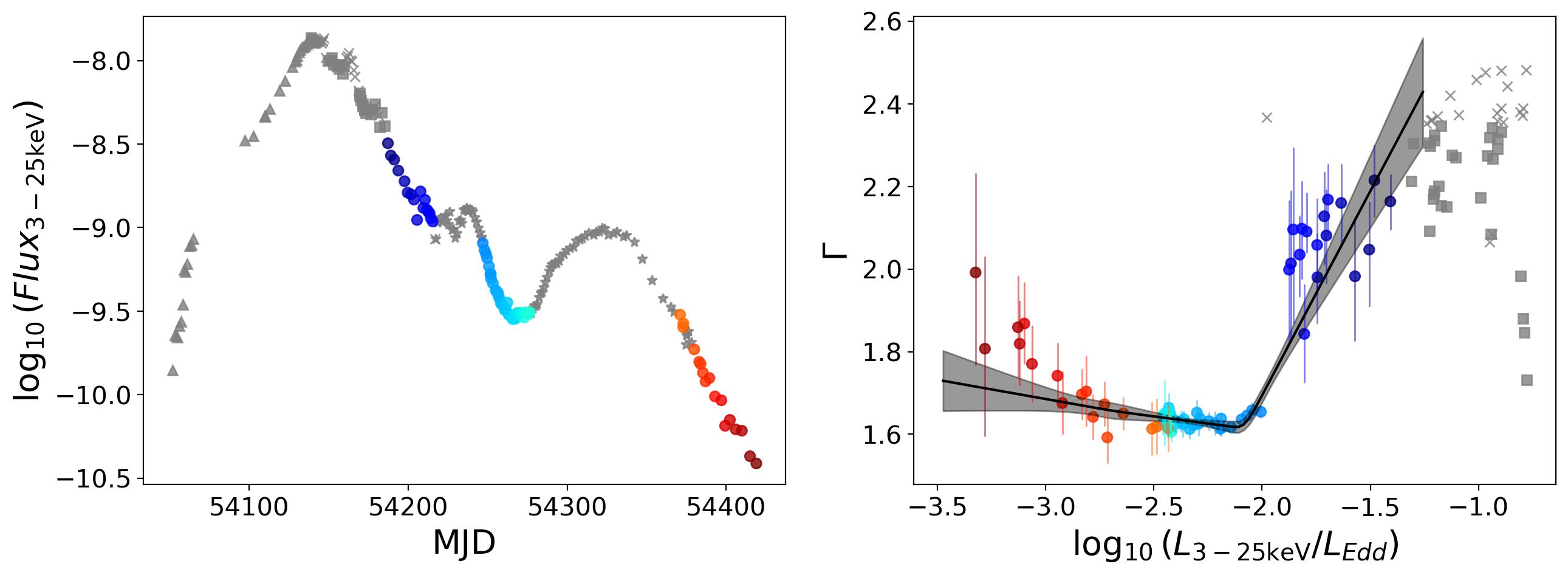}}
    \subfloat[GX 339-4 2010]{\includegraphics[width=0.28\textwidth]{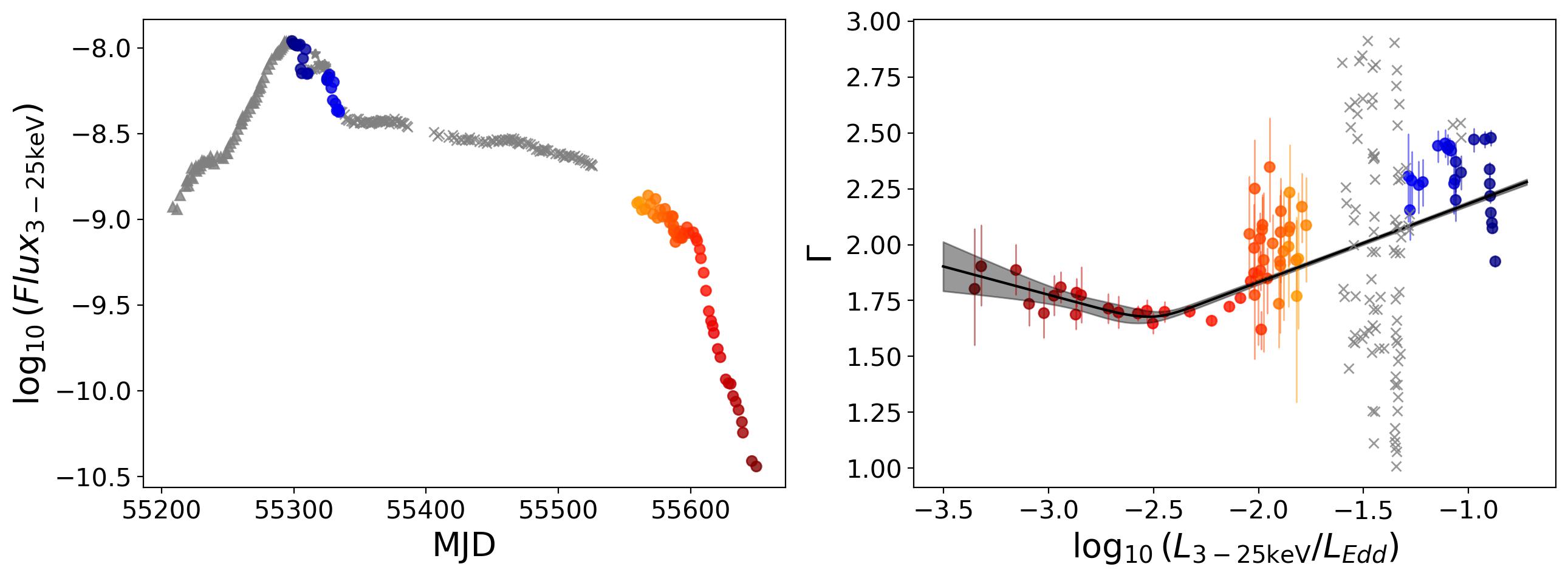}}
    \\
    \subfloat[XTE J1859+226]{\includegraphics[width=0.28\textwidth]{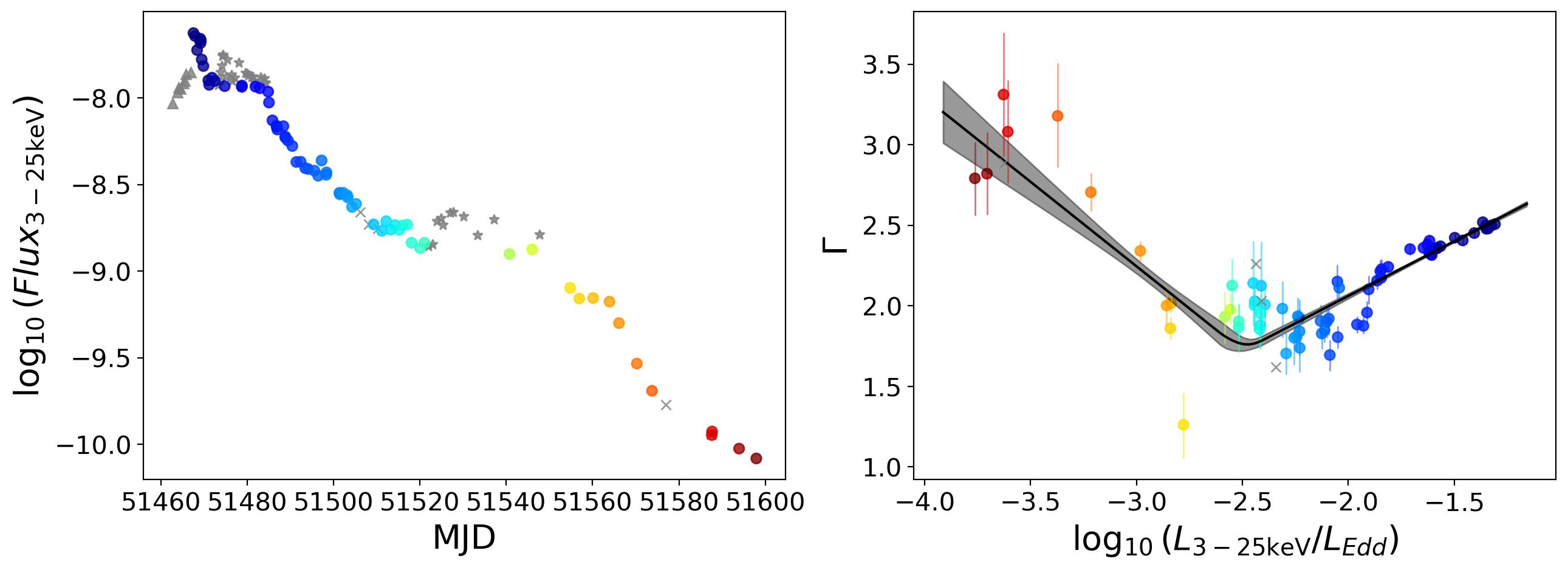}}
    \subfloat[GRO J1655-40]{\includegraphics[width=0.28\textwidth]{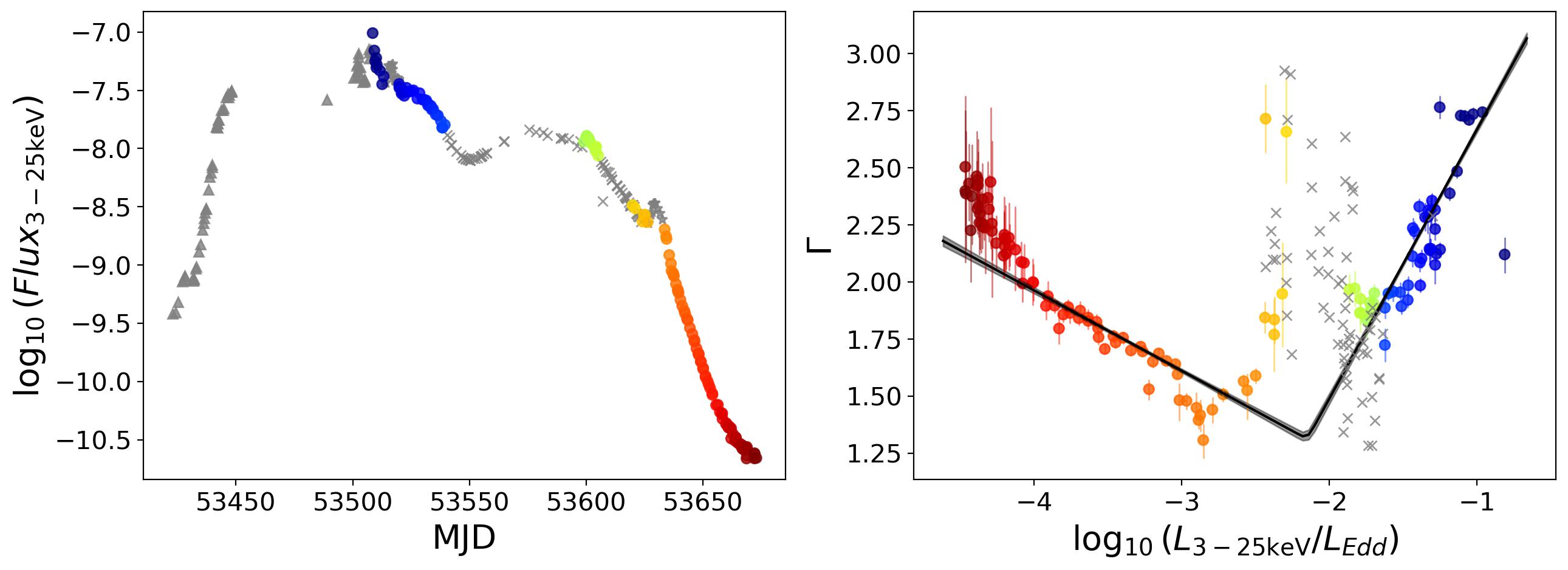}}
    \subfloat[XTE J1550-564 1998]{\includegraphics[width=0.28\textwidth]{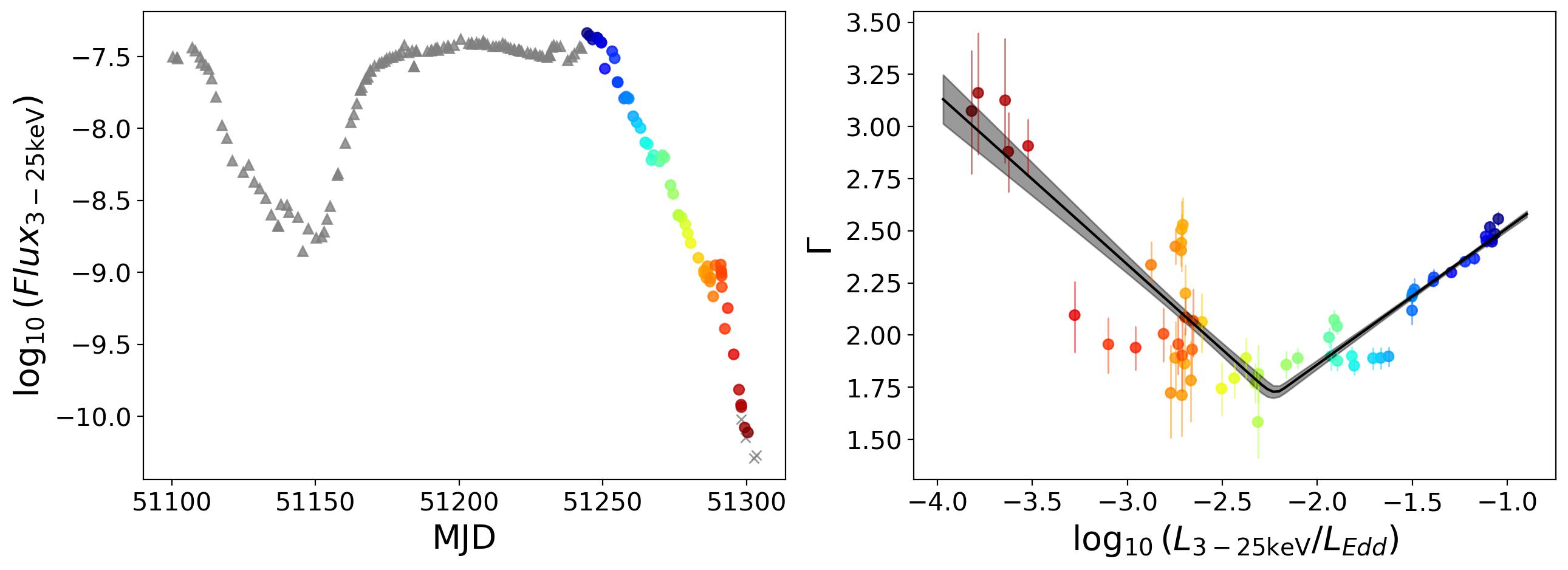}}
    \\
    \subfloat[XTE J1550-564 2000]{\includegraphics[width=0.28\textwidth]{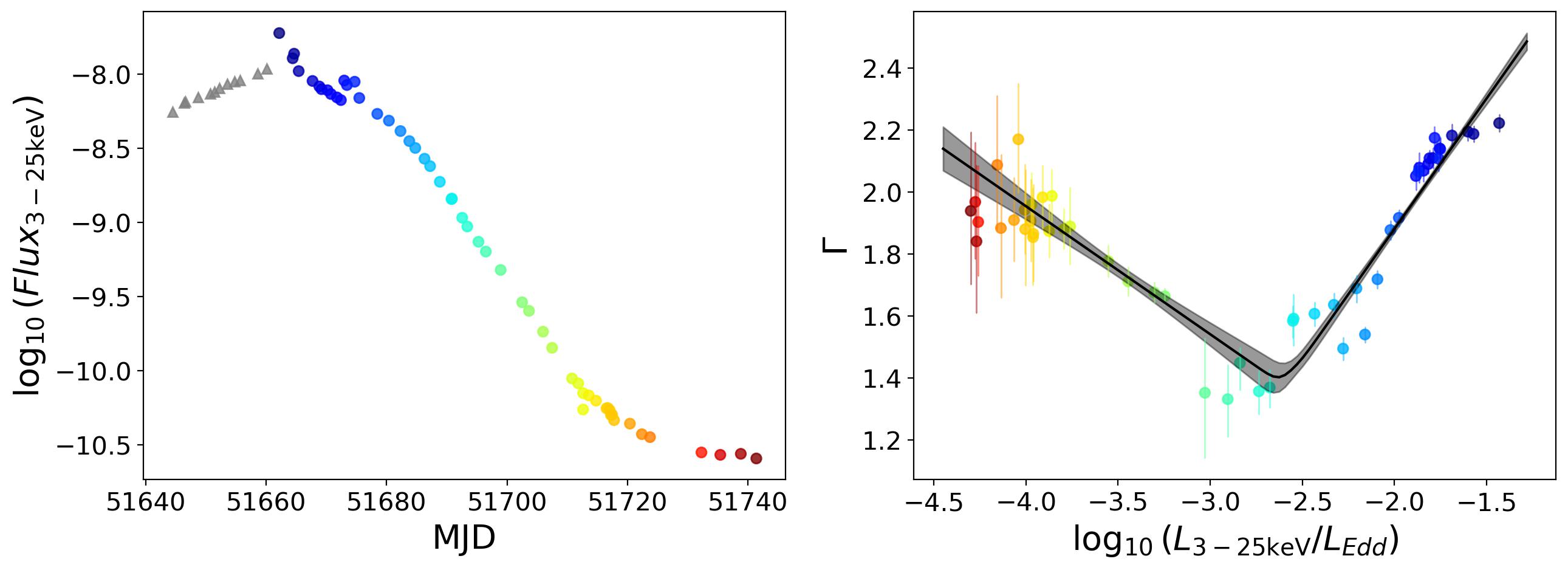}}
    \subfloat[H 1743-322 2003]{\includegraphics[width=0.28\textwidth]{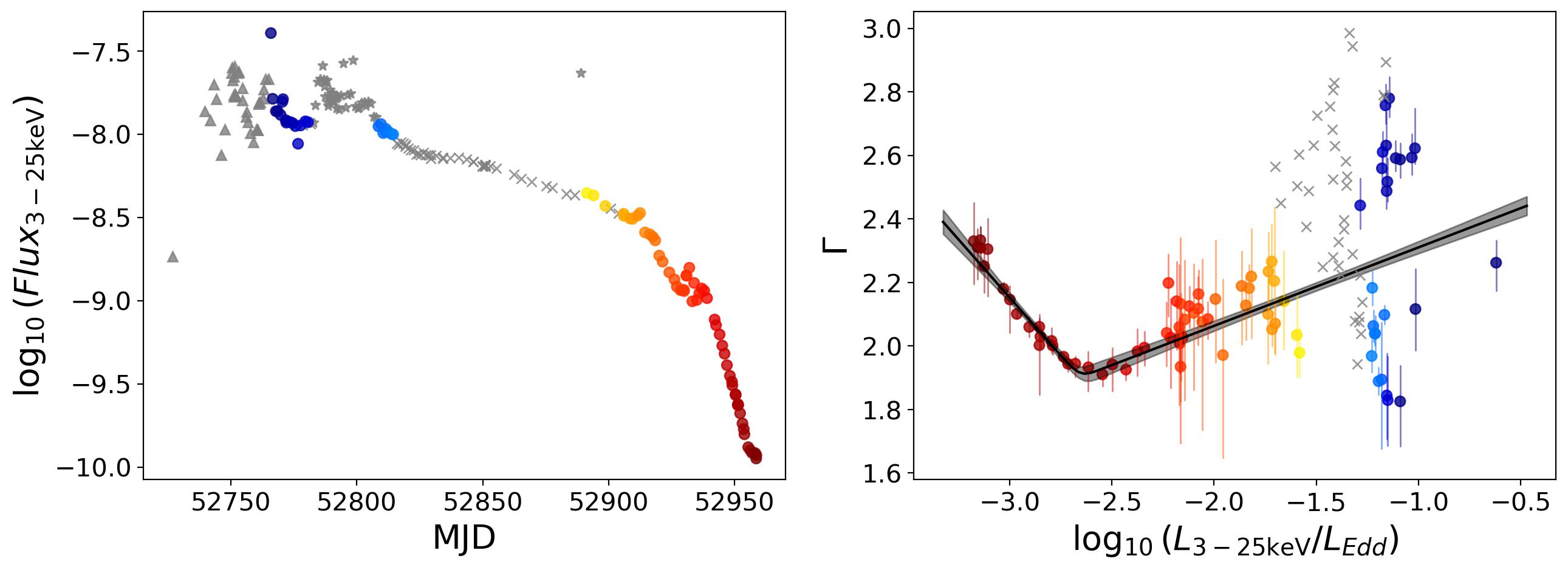}}
    \subfloat[H 1743-322 2004]{\includegraphics[width=0.28\textwidth]{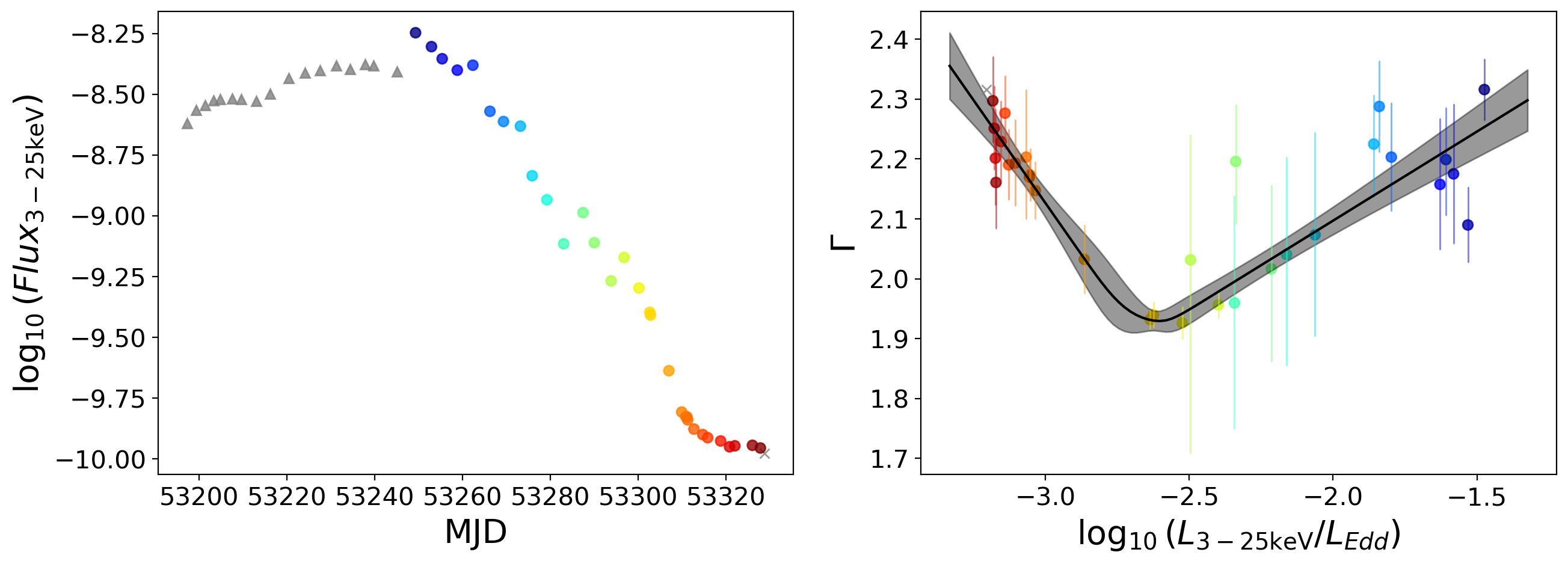}}
    \\
    \subfloat[H 1743-322 2005]{\includegraphics[width=0.28\textwidth]{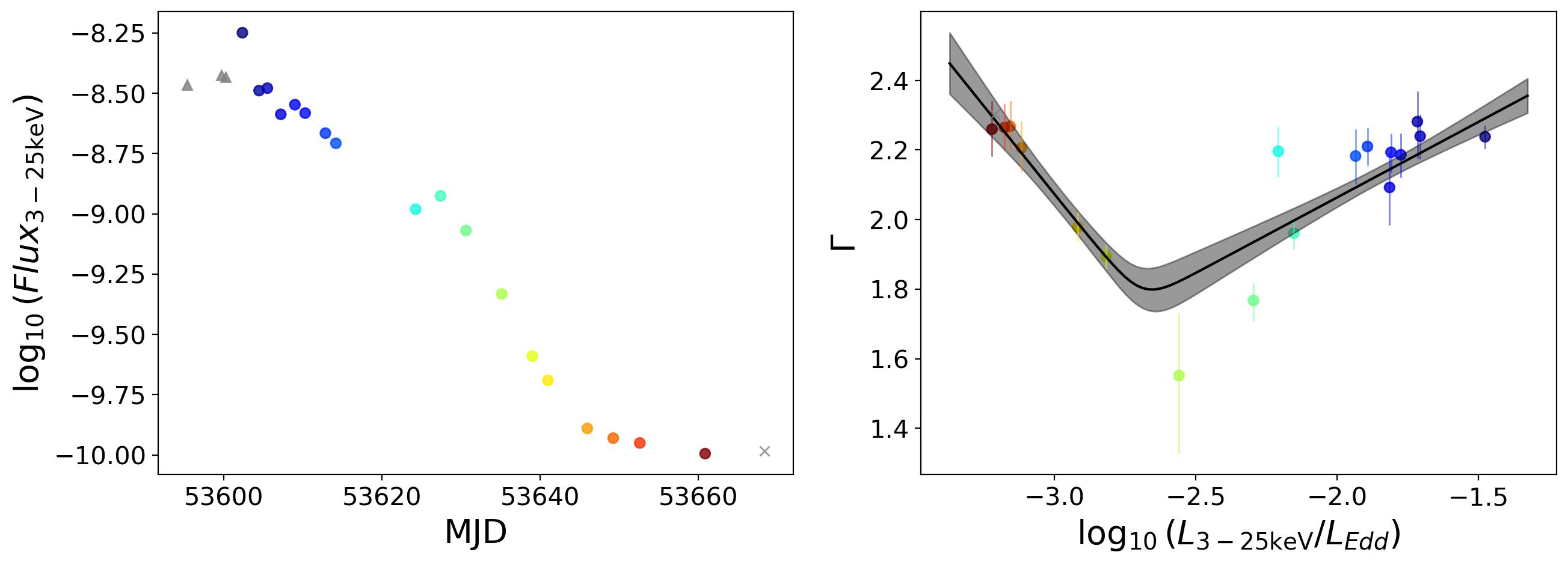}}
    \subfloat[H 1743-322 2009]{\includegraphics[width=0.28\textwidth]{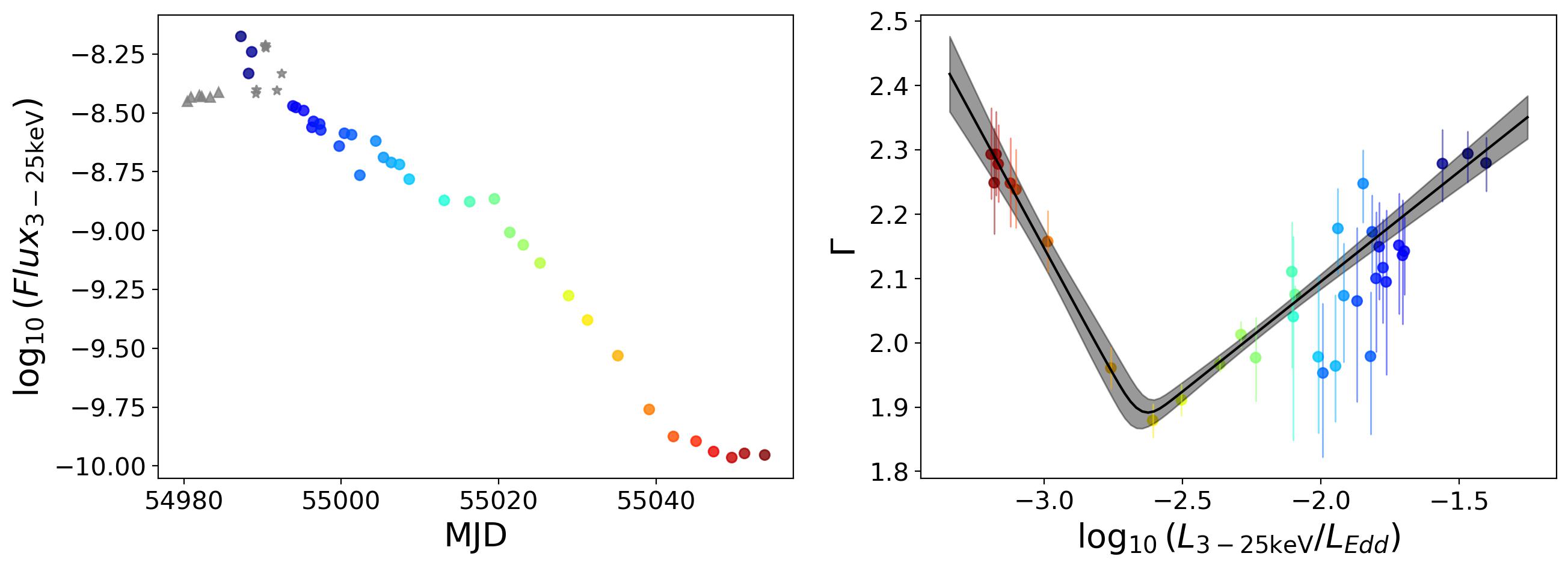}}
    \subfloat[MAXI J1659-152]{\includegraphics[width=0.28\textwidth]{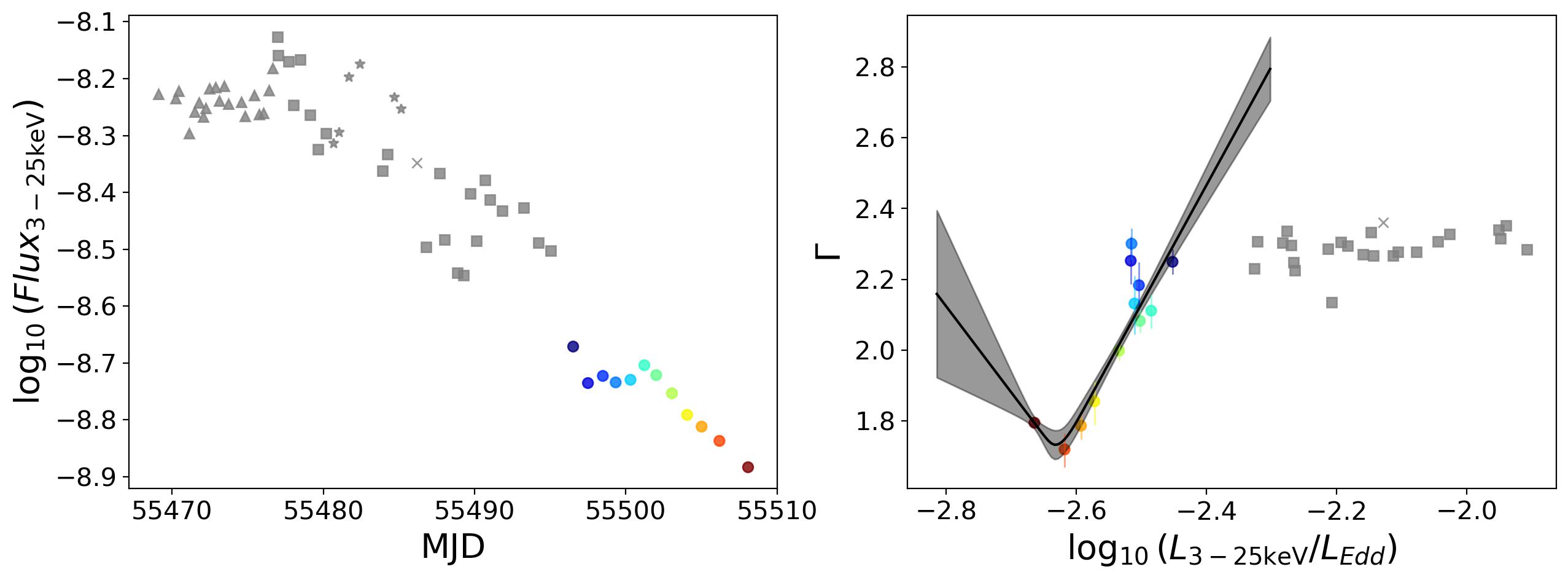}} 
    \\
\caption{Light curve (left figure of each subplot) and the fitting results of $\Gamma - \xi_X $ evolution diagram (right figure of each subplot) of each outburst of each source. In both left and right figures, the colored dots represent the data points that ultimately participated in the fitting of the $\Gamma - \xi_X $ evolution diagram and the color coding is time, the blue dots represent data points from earlier times and the red dots represent data points from later times; The gray triangle represent data points where the light curve is in the ascending stage; The gray X mark represents abnormal data points; The gray pentagrams indicate the data points that located within the ``flare", while the gray squares represent the data points where the $\Gamma$ hardly varies with the $\xi_X$. In the right figure of each subplot, the vertical lines of different colors passing through the colored points represent the error of the $\Gamma$ for that data point. The black curve represents the best-fitted curve, and the gray area indicates the $1\sigma$ confidence interval of the best-fitted curve.}
\label{f1}
\end{figure*}

Figure \ref{f1} presents the light curve and fitting results of the $\Gamma - \xi_X$ evolution diagram for each outburst of each source. The values and errors of the X-ray Eddington ratios at the cross points of the anticorrelation and positive correlation of $\Gamma - \xi_X $ evolution diagram for each fitting are presented in $ \rm Table $ \ref{t2}. From the fitting results, it can be seen that for the total 15 outbursts of 8 sources, the fitting effect of the anticorrelation and positive correlation of $\Gamma - \xi_X $ evolution diagram of most of the outbursts is good, except for the fitting result of the 2011 outburst of source MAXI J1659-152. The reason why the fitting result of $\Gamma - \xi_X$ evolution diagram fitting for the MAXI J1659-152 is not ideal is mainly due to the fact that after data screening, the number of data points involved in the fitting was few (especially the data points that involved in the anticorrelation fitting), which in turn led to a large error range for the fitting parameters.

\begin{deluxetable}{lccc}
\tablewidth{30pt}
\tablecaption{Positions of cross points of the anticorrelation and positive correlation of $\Gamma - \xi_X $ evolution diagram for each source \label{t2}}
\tablehead{
\colhead{Source} & \colhead{Outburst year} & \colhead{Inclination(degree)} & \colhead{$ \rm  log_{10}(\xi_{X_{cross \ point}}) $} \\
}
\startdata
4U 1543-47 & 2002 & $20.7\pm 1.5$ & $-2.32_{-0.07}^{+0.05}$ \\
\hline
XTE J1752-223 & 2009 & $35.0\pm 4.0$ & $-2.98_{-0.08}^{+0.07}$ \\
\hline
GX 339-4 & 2002 & $50.0\pm 10.0$ & $-2.25_{-0.03}^{+0.04}$  \\
         & 2004 & $50.0\pm 10.0$ & $-2.04_{-0.06}^{+0.04}$  \\
         & 2007 & $50.0\pm 10.0$ & $-2.09_{-0.03}^{+0.03}$  \\
         & 2010 & $50.0\pm 10.0$ & $-2.48_{-0.11}^{+0.09}$  \\
\hline
XTE J1859+226 & 1999 & $66.6\pm 4.3$ & $-2.49_{-0.08}^{+0.05}$  \\
\hline
GRO J1655-40 & 2005 & $70.2\pm 1.9$ & $-2.15_{-0.02}^{+0.02}$  \\
\hline
XTE J1550-564 & 1998 & $74.7\pm 3.8$ & $-2.22_{-0.04}^{+0.03}$  \\
              & 2000 & $74.7\pm 3.8$ & $-2.61_{-0.05}^{+0.08}$  \\
\hline
H 1743-322 & 2003 & $75.0\pm 3.0$ & $-2.65_{-0.04}^{+0.04}$  \\
           & 2004 & $75.0\pm 3.0$ & $-2.65_{-0.10}^{+0.08}$  \\
           & 2005 & $75.0\pm 3.0$ & $-2.69_{-0.07}^{+0.07}$  \\
           & 2009 & $75.0\pm 3.0$ & $-2.65_{-0.05}^{+0.05}$  \\
\hline
MAXI J1659-152 & 2011 & $75.0\pm 5.0$ & $-2.63_{-0.01}^{+0.02}$ 
\enddata
\tablecomments{}
\end{deluxetable}

\subsection{The linear fitting result of the inclination and \texorpdfstring{$\rm \xi_{X_{cross \ point}}$}{} }
\label{s3.2}
\begin{figure}
    \centering
    \plotone{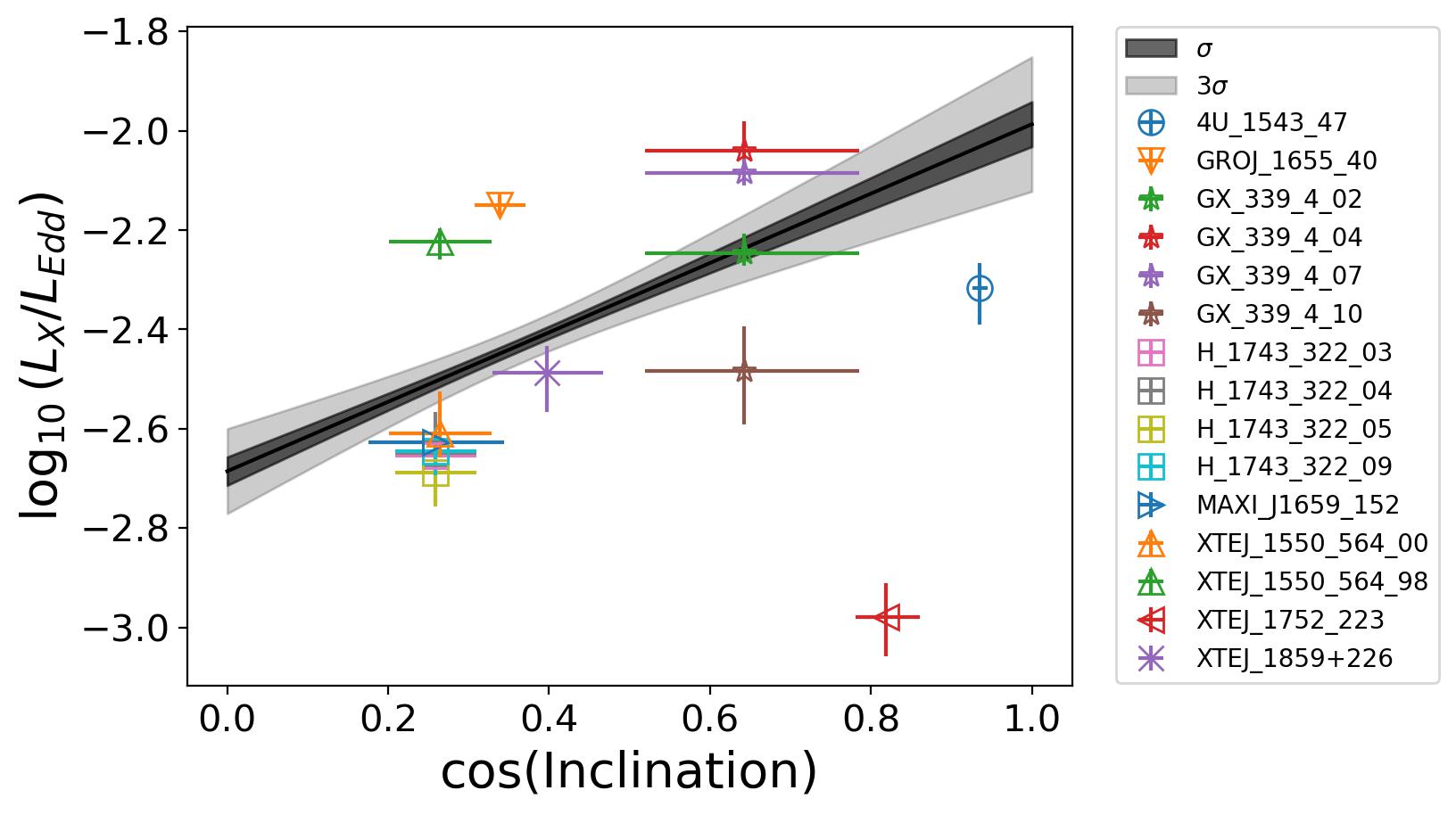}
    \caption{The result of fitting a linear relationship between and $\rm \xi_{X_{cross \ point}}$ of each source's individual outburst and cosine value of the inclination of this source. Different shapes of points represent the data of different sources. The black line is the best fitting line, and the black area is the $ 1\sigma $ confidence interval of the fitting line while the gray area is the $ 3\sigma $ confidence interval.}
    \label{f2}
\end{figure}

\begin{figure}
    \centering
    \plotone{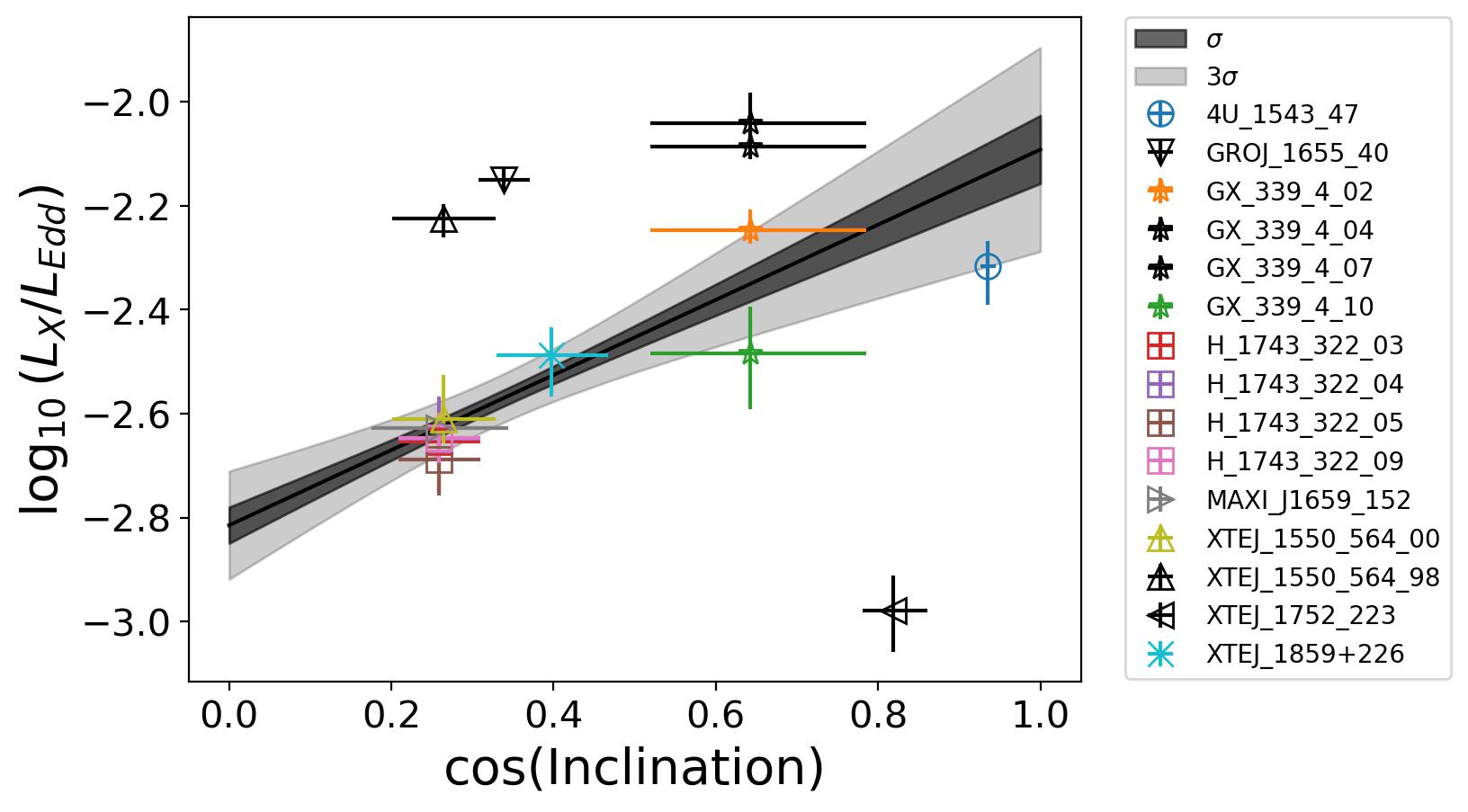}
    \caption{Similar to $\rm Figure$ \ref{f2}, however, the outburst of the source GX 339-4 in 2004 and 2007, the outbursts of the source GRO J1655-40 in 2005, the outburst of the source XTE J1752-223 in 2009 and  the outburst of the source XTE J1550-564 in 1998 do not participate in the linear relationship fitting. Different shapes of black dots represent the points that do not participate in the fitting.}
    \label{f3}
    \end{figure}

$\rm Figure $ \ref{f2} shows the result of the linear relationship fitting between the inclination and $ \rm \xi_{X_{cross \ point}} $ of all the sources. From the final displayed fitting diagram, we can see that for the sources with multiple outbursts, such as H 1743-322, their data points are basically in the same area on the result diagram of the fitted linear relationship. Meanwhile, we can observe that for most sources, there seems to be a positive linear relationship between the inclination and $\rm \xi_{X_{cross \ point}}$. However, for some outbursts, the $\rm \xi_{X_{cross \ point}}$ deviates significantly from the best-fitted straight line. This part of the outbursts includes the 2004 and 2007 outbursts of source GX 339-4, the 2005 outburst of source GRO J1655-40, the 2009 outburst of source XTE J1752-223, and the 1998 outburst of source XTE J1550-564. The detailed discussion of these five outbursts will be given in Section \ref{s4.4}. To obtain more accurate linear fitting results, we have provided the linear fitting results of these five outbursts that do not participate in the linear fitting in $\rm Figure $ \ref{f3}. It can be seen that there is a relatively obvious positive linear correlation between the inclination of the remaining outbursts and the $\rm \xi_{X_{cross \ point}}$. Quantitatively: $ \rm log_{10}(\xi_{X_{cross \ point}}) = 0.72^{+0.09}_{-0.09}\cos{\theta} - 2.81^{+0.03}_{-0.03} $, where $ \rm \theta $ is the inclination of BH XRBs.

\section{Discussion} \label{s4}

\subsection{The fitting effect of the photon spectral index and the X-Ray Eddington ratio evolution diagram and linear relationship} 
\label{s4.1}

First, we use the four outbursts of H 1743-322 to illustrate the fitting results of the $\Gamma - \xi_X$ evolution diagram. Figure \ref{f1}(g)--(j) show the fitting results for these four outbursts. For all four outbursts, both the anticorrelation and positive correlation are well fitted. Moreover, as shown in Table \ref{t2}, the cross point positions ($\xi_{\mathrm{X}_{\mathrm{cross\ point}}}$) for the four outbursts are approximately the same ($ \rm -2.69<  \xi_{X_{cross \ point}} < -2.65 $). This consistency is expected if the underlying physical mechanism governing the $\Gamma$-$\xi_X$ relationship is the same for a given source, and if the inclination, which is a geometric property of the system, does not change significantly over time. Our results are therefore consistent with the geometric interpretation proposed by \citet{Xu2015} for AGN, and suggest that similar projection effects may be at work in BH XRBs.

For the linear relationship fitting, as seen in Table \ref{t2}, the $\xi_{X_{cross \ point}}$ values for the four outbursts of H~1743-322 are nearly identical. Consequently, in Figure \ref{f2} and Figure \ref{f3}, the data points for these four outbursts cluster in the same region. This further supports the physical picture we propose, based on the work of \citet{Xu2015}: for a given BH XRB, the inclination is essentially constant, and therefore the observed $\xi_{X_{cross \ point}}$ should remain approximately the same across multiple outbursts. Moreover, most data points in Figure \ref{f3} lie close to a straight line with a positive slope, indicating a clear positive correlation between $\xi_{X_{cross \ point}}$ and $\cos\theta$ across different BH XRBs. Quantitatively, we find $ \rm log_{10}(\xi_{X_{cross \ point}}) = 0.72^{+0.09}_{-0.09}\cos{\theta} - 2.81^{+0.03}_{-0.03} $, where $ \rm \theta $ is the inclination of BH XRBs.

\subsection{Physical origin of the cross point: A Unified Disk-corona-ADAF Model}
\label{s4.2}

We further explore the physical mechanism that may give rise to the cross point in the $\Gamma$-$\xi_X$ evolution diagram. The observed photon spectral index $\Gamma$ reflects the Comptonization process in the corona, where the electron energy distribution is governed by the balance between heating and cooling \citep{1979rpa..book.....R,Beloborodov1999,Nttil2024}. The position of the cross point may therefore mark a transition in the dominant cooling mechanism of the corona \citep{Qiao2013}.

A theoretical framework that naturally accounts for such a transition is the disk-corona evaporation/condensation model. \citet{Cao2009} constructed an accretion disk-corona model for AGN (\citet{Xu2015} also used the same model) and found that the hard X-ray spectral index correlates positively with the Eddington ratio at high accretion rates, consistent with observations. However, the model fails to reproduce the very hard X-ray spectra observed in sources accreting at low rates (e.g., $\Gamma \sim 1$ at $\rm L_{bol}/L_{Edd} \sim 0.01$), leading \citet{Cao2009} to suggest that the accretion mode may transition to an ADAF + outer disk-corona system at low accretion rates. This picture was later developed into a quantitative model for BH XRBs by \citet{Qiao2013}, who explicitly demonstrated that the $\Gamma$-$\xi_X$ correlation in BH XRBs can be explained by a two-phase accretion flow: At high accretion rates ($\xi_X \gtrsim 0.01$), a fraction of coronal gas condenses into the disk, sustaining an inner disk. The X-ray emission is dominated by Comptonization of disk soft photons in the corona, yielding a positive $\Gamma$-$\xi_X$ correlation. At low accretion rates ($\xi_X \lesssim 0.01$), evaporation becomes dominant, the inner disk vanishes completely, and the inner region becomes an ADAF. The X-ray emission arises from Comptonization of the ADAF's own synchrotron and bremsstrahlung photons, yielding an anticorrelation. This model successfully explains the complete shape of the observed $\Gamma$-$\xi_X$ relationship in BH XRBs. It also naturally explains why the same correlation is observed in both AGN and BH XRBs, they share the same ``disk + corona + ADAF'' accretion structure, differing only in scale \citep{Yang2014,Yuan2014,Ruan2019}.

\subsection{Inclination effect: geometrical projection superimposed on the accretion structure}
\label{s4.3}

While the \citet{Qiao2013} model successfully explains the shape of the $\Gamma$-$\xi_X$ relation, it treats the system as one-dimensional along the accretion disk plane and does not consider the viewing geometry. Our results demonstrate that the position of the cross point, the critical $\xi_X$ at which the transition from ADAF-dominated to disk-dominated accretion occurs, is systematically modulated by the inclination. This can be naturally understood as a geometric projection effect superimposed on the two-phase accretion structure.

In the \citet{Qiao2013} picture, the intrinsic transition of the two correlations occurs at a critical accretion rate $\xi_X \sim 0.01$ in the comoving frame of the accretion flow. This intrinsic transition rate is determined by the accretion physics and is independent of the inclination. However, we think that the observed transition point $\xi_{X_{cross\ point}}$ will be modulated by the inclination through three effects.

The first effect is the projection effect due to the inclination mentioned in Section \ref{s4.1}. This is a purely geometric effect: the observed flux is reduced by a factor of $\cos \theta$, independent of photon energy. Although the intrinsic $\xi_{X_{cross\ point}}$ is independent of the inclination, sources with high inclination will observe lower X-ray luminosity due to this projection effect, while sources with low inclination will observe greater X-ray
luminosity.

The second effect is the anisotropy of radiation that may exist in the disk-corona model of BH XRBs. \citet{Xu2015} found that the disk-corona model used for AGN radiation exhibits anisotropic radiation in terms of the X-ray specific intensity: it is nearly independent of viewing angle at small inclinations, but becomes strongly sensitive to viewing angle at large inclinations. \citet{Zhang2019} further confirmed that for a corona located above the accretion disk, the emergent spectrum is not isotropic, with harder and less luminous spectra toward observers at lower inclinations, owing to the anisotropic illumination of seed photons. These two results concern different physical quantities: the former focuses on the angle-dependent nature of a specific intensity, while the latter examines the inclination-dependent nature of the observed spectral shape and brightness. Therefore, they are rather complementary manifestations of the same fundamental anisotropy phenomenon. We can therefore reasonably infer that the disk-corona model of BH XRBs may also exhibit inclination-dependent radiation anisotropy. In the picture of \citet{Qiao2013}, the accretion flow undergoes a transition from the external ADAF to the internal disk-corona system along the radial direction, while the vertical structure adopts a simplified self-similar treatment. This asymmetry in the treatment of the radial and vertical structures may modulate the anisotropic radiation, although the precise direction and magnitude of the effect require dedicated radiative transfer calculations.

The third effect is the possible absorption of soft X-ray photons by the cold disk at the outer part of high-inclination BH XRBs. \citet{Narayan2005} found through statistics that among known BH XRBs, no system has an inclination exceeding $75^{\circ}$, and none of the 20 samples exhibits an eclipse. This strongly suggests that high-inclination systems are obscured. \citet{Narayan2005} proposed that the possible source of the obscuration is disk flaring, that is, the outer disk rises in the vertical direction, forming a shield of the central radiation source along the high-inclination line of sight. In the \citet{Qiao2013} model, the outer region of the accretion flow is composed of a truncated cold disk, and the corona is cooled by inverse Compton scattering of soft photons from the cold disk, with radiative coupling and mass exchange between the cold disk and the corona \citep{Liu2007,Qiao2013}. Combining the results of \citet{Narayan2005} and the model of \citet{Qiao2013}, we can reasonably infer that for high-inclination BH XRBs, the cold disk may exhibit the disk flaring and will absorb soft photons produced by the corona, thereby reducing the observed X-ray luminosity and finally affecting $\xi_{X_{cross\ point}}$ to be lower. However, this effect also requires quantitative calculation to be compared with observations.

Crucially, MONK simulations have directly demonstrated the impact of inclination on coronal spectra \citep{Fan2025}. In the ``sandwich'' (disk-like) corona geometry, the path length of a photon through the corona increases with inclination as $1/\cos \theta$, so that the effective optical depth experienced by the escaping photons is larger at higher inclination. This results in systematically lower $\Gamma$ and disk fraction at higher inclination. Although the MONK simulations use a different simplified model from our observational analysis, their physical conclusion that inclination modulates the coronal spectrum by altering the effective optical depth is qualitatively consistent with our observational results. Our findings can therefore be viewed as an observational validation of the inclination effects revealed by the MONK simulations.

\begin{figure}
    \centering
    \plotone{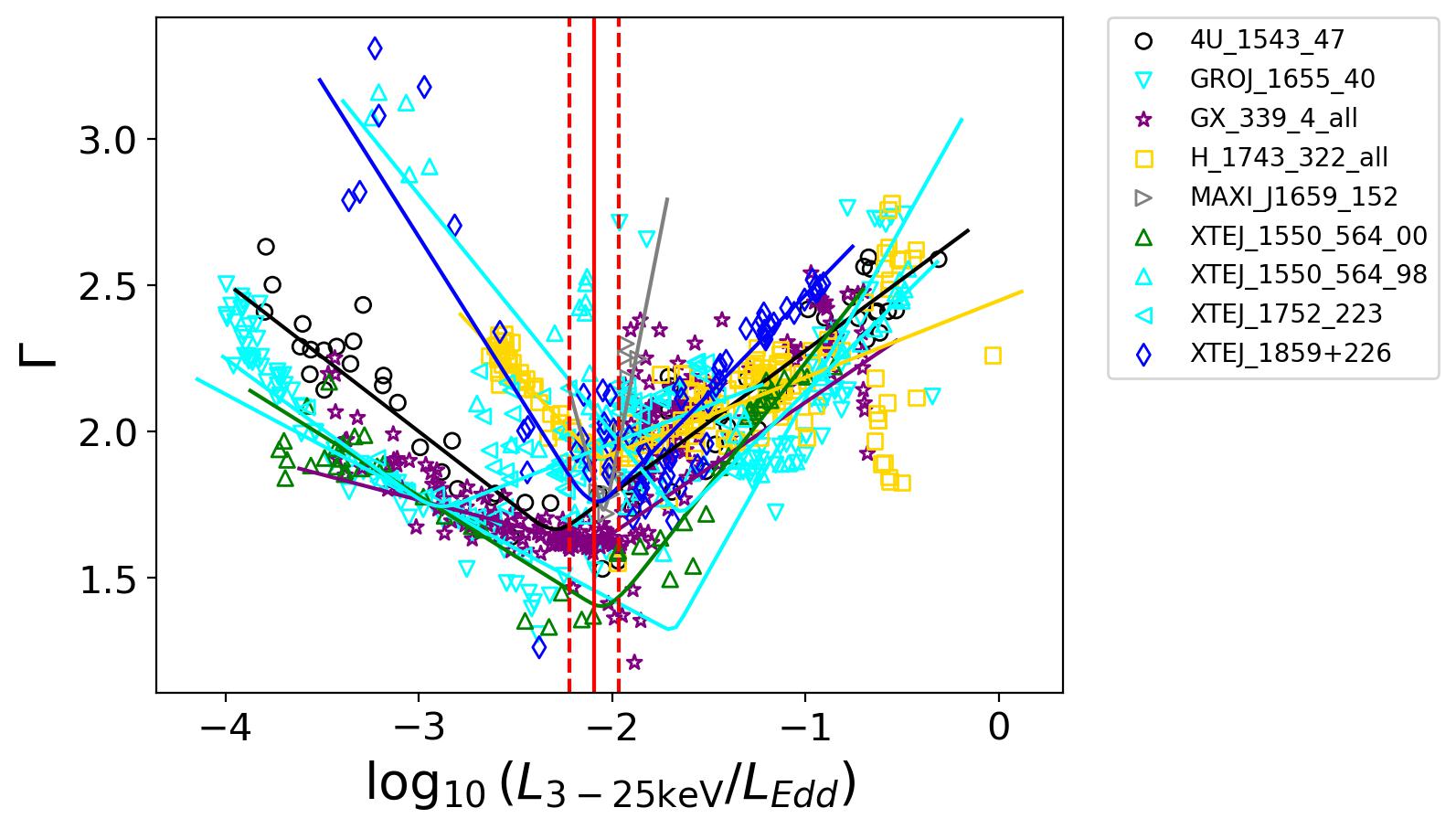}
    \caption{Normalized $\Gamma-\xi_X$ relationship plot after normalizing $\xi_X$ using the $cos \theta$ of each source ($log10(\xi_X) - log10(cos\theta)$). Different colors and shapes represent data from different sources and their best-fit results. The cyan points and lines indicate the data and best-fit results for sources deviating from the linear relationship between $cos\theta$ and $\xi_{X_{cross \ point}}$. The red solid line and dashed line represent, respectively, the $\xi_{X_{cross \ point}}$ value at $\theta$ = 0 and its $1\sigma$ uncertainty, derived from the linear fit of the $\cos\theta$ versus $\xi_{X_{cross \ point}}$ relationship.}
    \label{f4}
\end{figure}

To quantitatively test whether the observed diversity in the $\Gamma$-$\xi_X$ relation across sources can be attributed to inclination, we plot all data points with the X-axis normalized by $\cos\theta$ (Figure \ref{f4}). After this simple geometric correction, the data points from different sources exhibit significantly reduced scatter and converge toward a common trend. This provides strong empirical support for the interpretation that the apparent differences in the spectral evolution among BH XRBs are primarily driven by the geometric projection effect, while the underlying accretion physics described by the \citet{Qiao2013} model is broadly consistent across sources. We note, however, that this conclusion is based on a limited sample and that additional effects, such as the cold disk absorption discussed above, may also contribute to the residual scatter.

Thus, the observed positive correlation between $\xi_{X_{cross\ point}}$ and $\cos\theta$ arises from a combination of two factors: the intrinsic transition in coronal cooling physics that occurs at $\xi_X \sim 0.01$ (the disk-corona + ADAF two-phase structure), and the geometrical projection of the disk-like corona, which shifts the apparent transition point to higher or lower $\xi_X$ depending on the line-of-sight inclination.

\subsection{Outbursts of sources deviating from the best-fit linear relationship}
\label{s4.4}

Although the most sample shows a clear positive linear correlation between $ \rm \xi_{X_{cross \ point}} $ and $\cos \theta$, as can be seen from Figure \ref{f3}, there are five outbursts whose $ \rm \xi_{X_{cross \ point}} $ values deviate from our best-fit line. These five outbursts are the 2004 and 2007 outbursts of source GX 339-4, the 2005 outburst of source GRO J1655-40, the 1998 outburst of XTE J1550-564, and the 2009 outburst of source XTE J1752-223. Moreover, the $ \rm \xi_{X_{cross \ point}} $ values of the 2004 and 2007 outbursts of source GX 339-4, the 2005 outburst of GRO J1655-40 and the 1998 outburst of XTE J1550-564 are greater than the predicted values of the best-fitted line, while the $ \rm \xi_{X_{cross \ point}} $ value of the 2009 outburst of XTE J1752-223 is less than the predicted value of the best-fitted line. We will discuss the reasons for the $ \rm \xi_{X_{cross \ point}} $ deviations of these five outbursts from the best-fitted line separately below.

\subsubsection{The 2004 and 2007 outbursts of source GX 339-4}
\begin{figure*}[t] 
    \centering
    \begin{subfigure}{0.64\textwidth}
        \centering
        \includegraphics[width=\textwidth]{GX_339_4_04.jpg}
    \end{subfigure}
    \hfill 
    \begin{subfigure}[b]{0.34\textwidth}
        \centering
        \includegraphics[width=\textwidth]{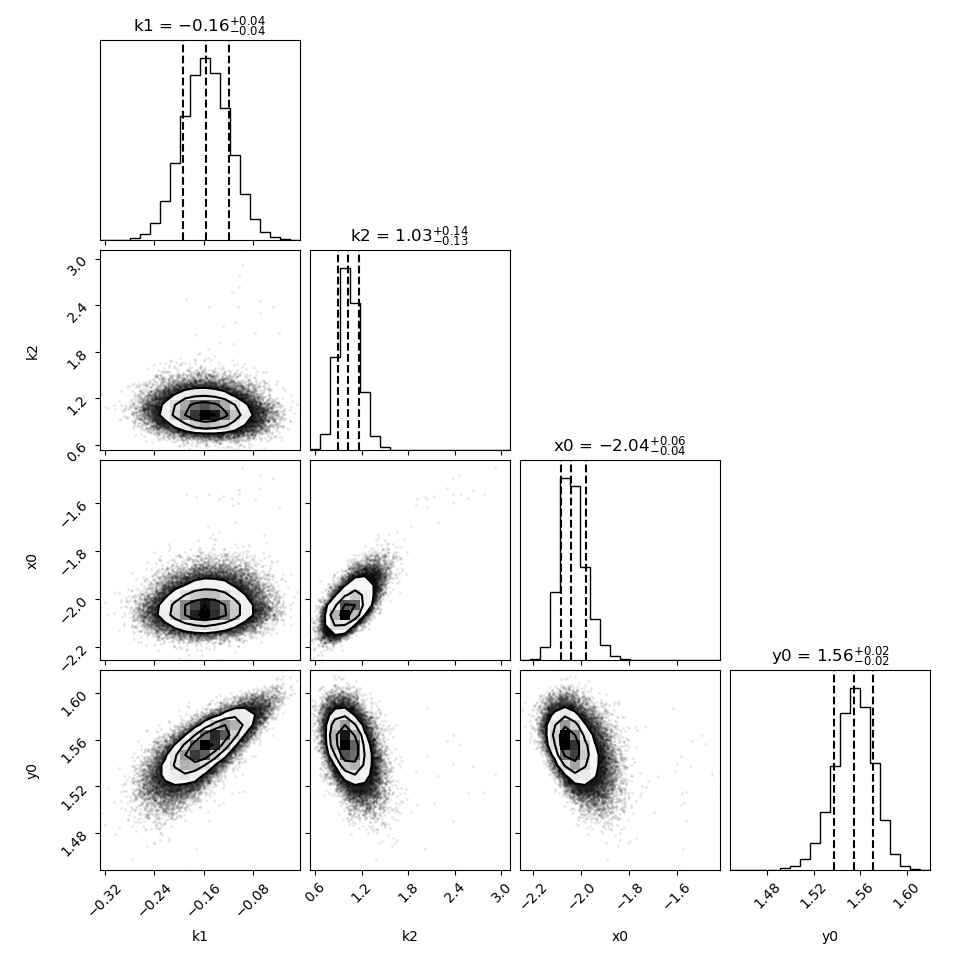}
    \end{subfigure}
   \\
    \makebox[\textwidth]{\footnotesize GX 339-4 2004}
    \\[4ex] 
    \begin{subfigure}{0.64\textwidth}
        \centering
        \includegraphics[width=\textwidth]{GX_339_4_07.jpg}
    \end{subfigure}
    \hfill 
    \begin{subfigure}[b]{0.34\textwidth}
        \centering
        \includegraphics[width=\textwidth]{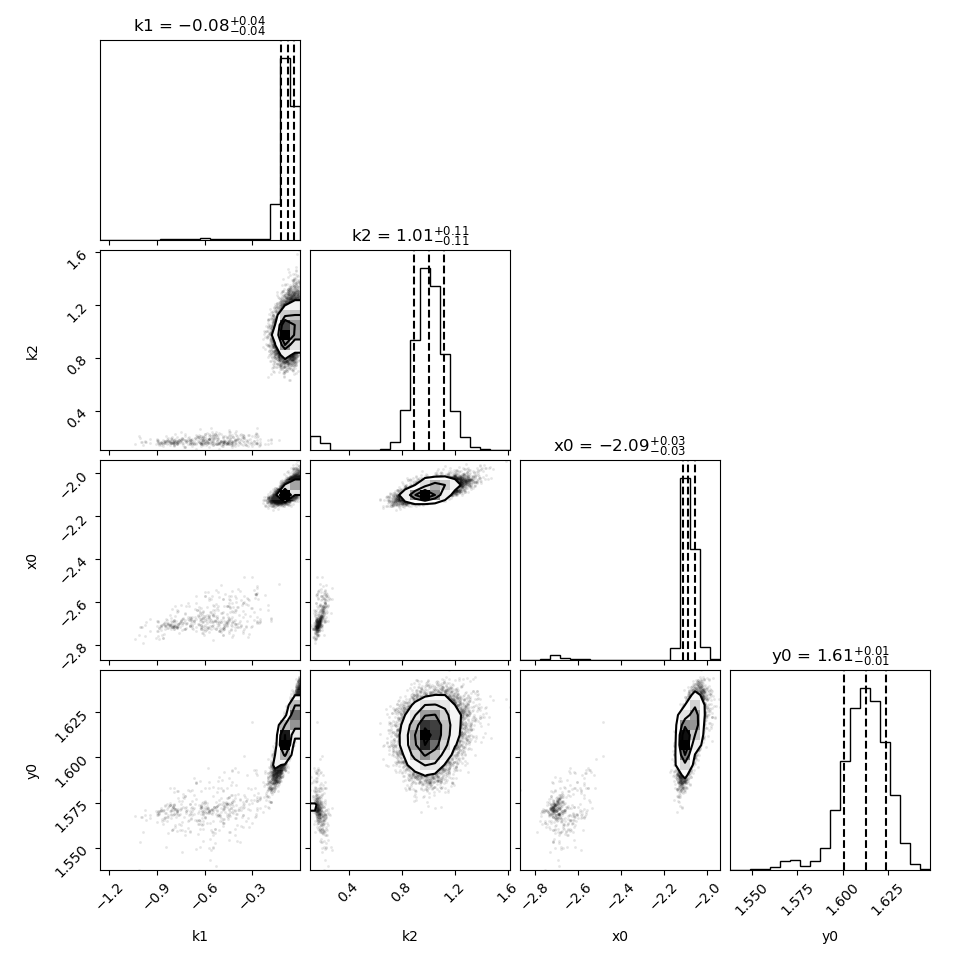}
    \end{subfigure}
   \\
    \makebox[\textwidth]{\footnotesize GX 339-4 2007}
    \caption{Data processing results of the 2004 and 2007 outbursts of source GX 339-4. For each outburst, the left figure is the light curve, the middle figure is the $\Gamma-\xi_X$ plot and its fitting result, and the right figure is the corner of the fitting result}
    \label{f5}
\end{figure*}

GX 339-4 stands out as one of the most important target sources for studying the physics of BH accretion due to its frequent outbursts and the most complete state evolution \citep{Nandi2012,Bhowmick2021}. Figure \ref{f5} shows the fitting result of the 2004 and 2007 outbursts of source GX 339-4. The reason why the fitting results of $\Gamma - \xi_X$ evolution diagram deviating from the best-fit linear relationship for the 2002 and 2004 outbursts of the source GX 339-4 is that, on the one hand, the variation of $\Gamma$ with $\xi_X$ in these two outbursts is relatively weak in the cross area of the $\Gamma - \xi_X$ relationship diagram. Moreover, the data points showing this situation for the 2002 and 2010 outbursts of source GX 339-4 were relatively few. This area will cause the cross point to move to a region far from the minimum square result after the MCMC method; on the other hand, after data screening, the number of data points participating in the fit for a single outburst is relatively small, which will also, to some extent, affect the result of the MCMC method. According to the discussion in Section \ref{s4.1}: for the same BH XRBs, we generally think that their inclination is almost unchanged, so the $\Gamma - \xi_X$ evolution diagrams and the value of $ \xi_{X_{cross \ point}} $ should be approximately the same. Therefore, in response to the above situation, for the sources GX 339-4 and H 1743-322 with multiple outbursts, we can place all the filtered data points of the descending stages of each outburst on one diagram and fit them together to obtain a more accurate cross point coordinate (For the source XTE J1550-564, there are fundamentally different spectral evolution patterns in its 1998 and 2000 outbursts \citep{Titarchuk2010}. Therefore, it is not appropriate to combine the data from the two outbursts for fitting. The detailed discussion on this matter will be presented in Section \ref{s4.4.3}).  

\begin{figure*}[t] 
    \centering
    \begin{subfigure}{0.64\textwidth}
        \centering
        \includegraphics[width=\textwidth]{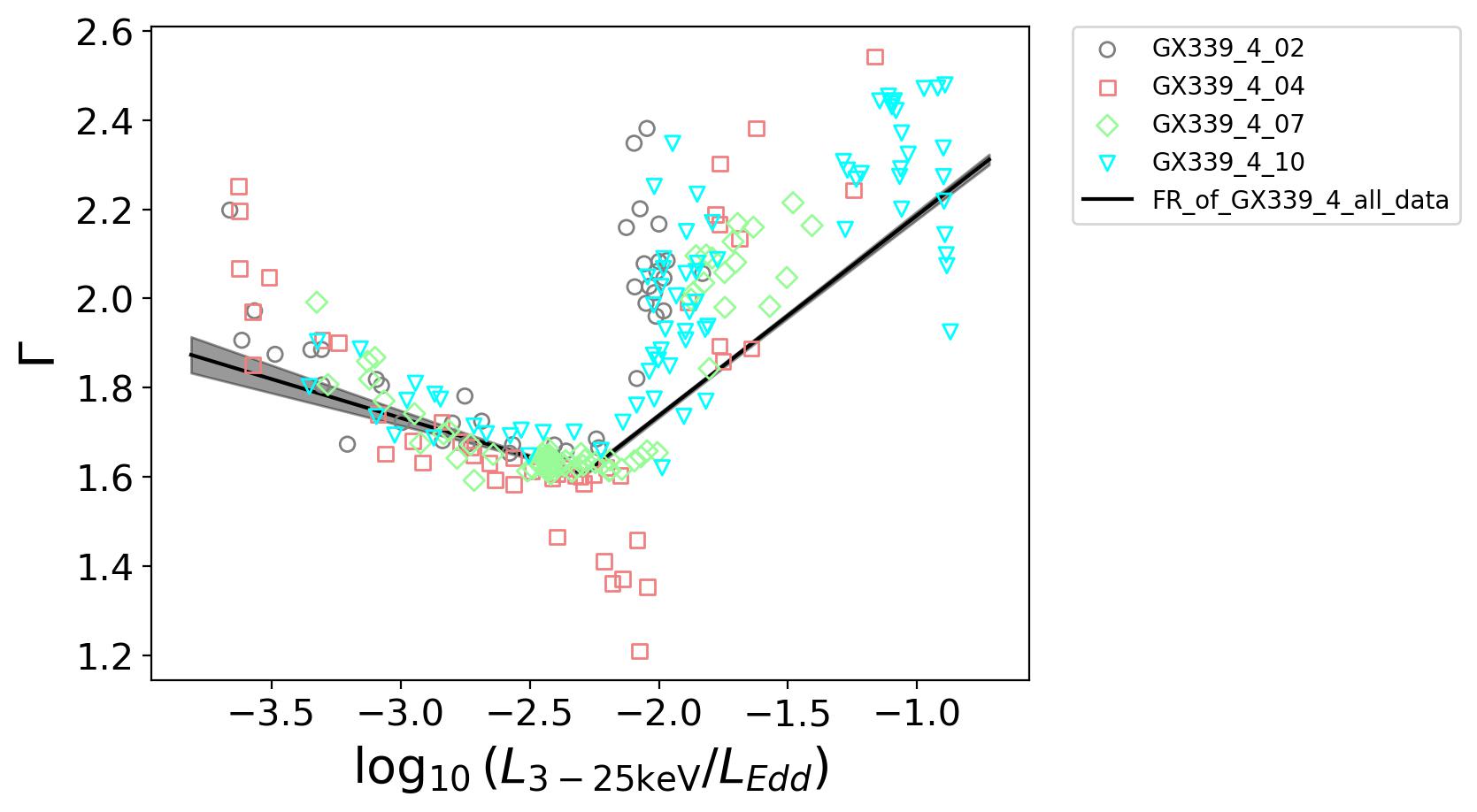}
    \end{subfigure}
    \hfill 
    \begin{subfigure}[b]{0.34\textwidth}
        \centering
        \includegraphics[width=\textwidth]{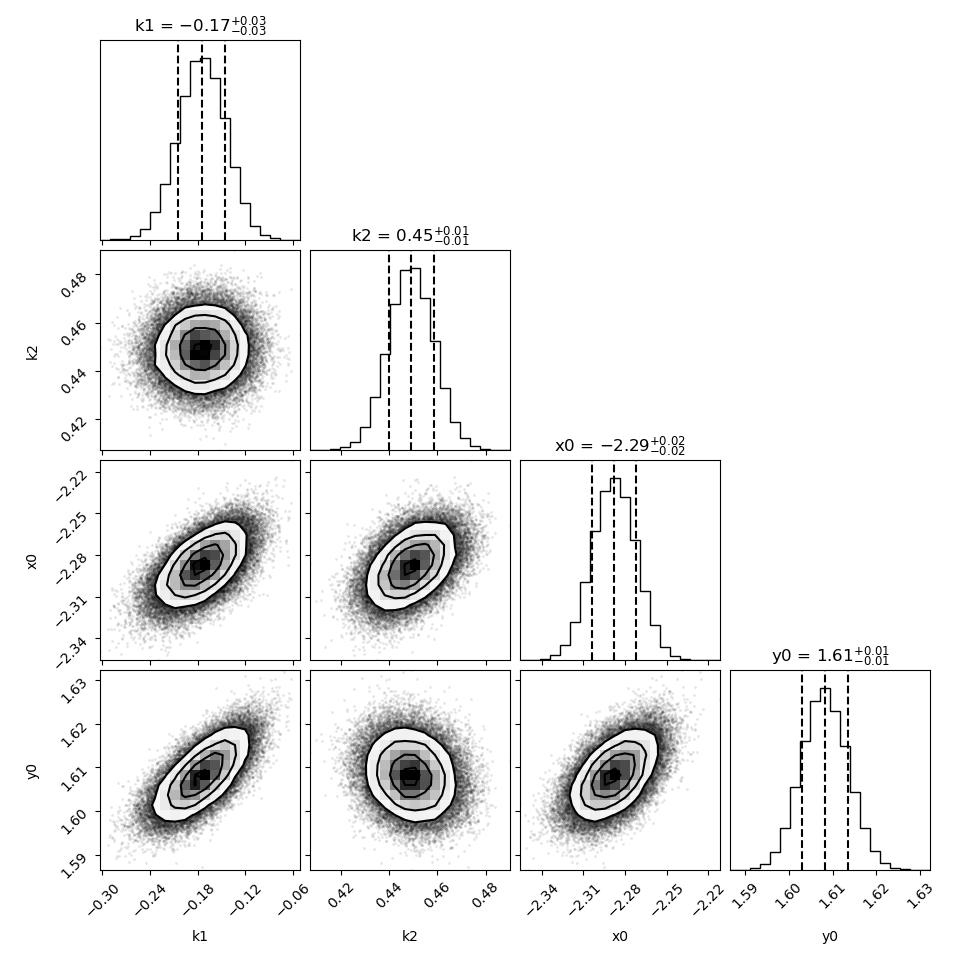}
    \end{subfigure}
   \\
    \makebox[\textwidth]{\footnotesize(1) GX 339-4 all}
    \\[5ex] 
    \begin{subfigure}{0.64\textwidth}
        \centering
        \includegraphics[width=\textwidth]{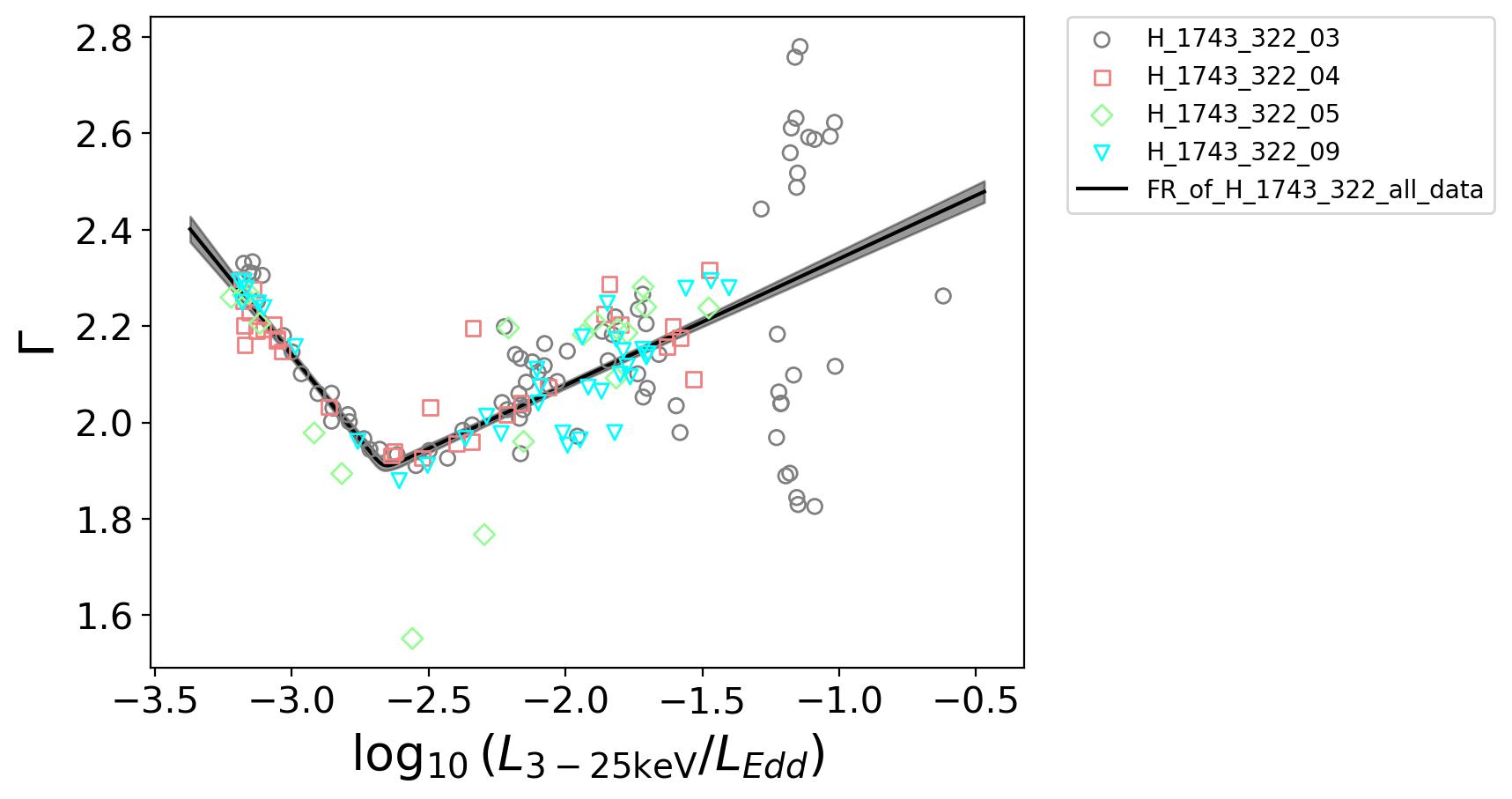}
    \end{subfigure}
    \hfill 
    \begin{subfigure}[b]{0.34\textwidth}
        \centering
        \includegraphics[width=\textwidth]{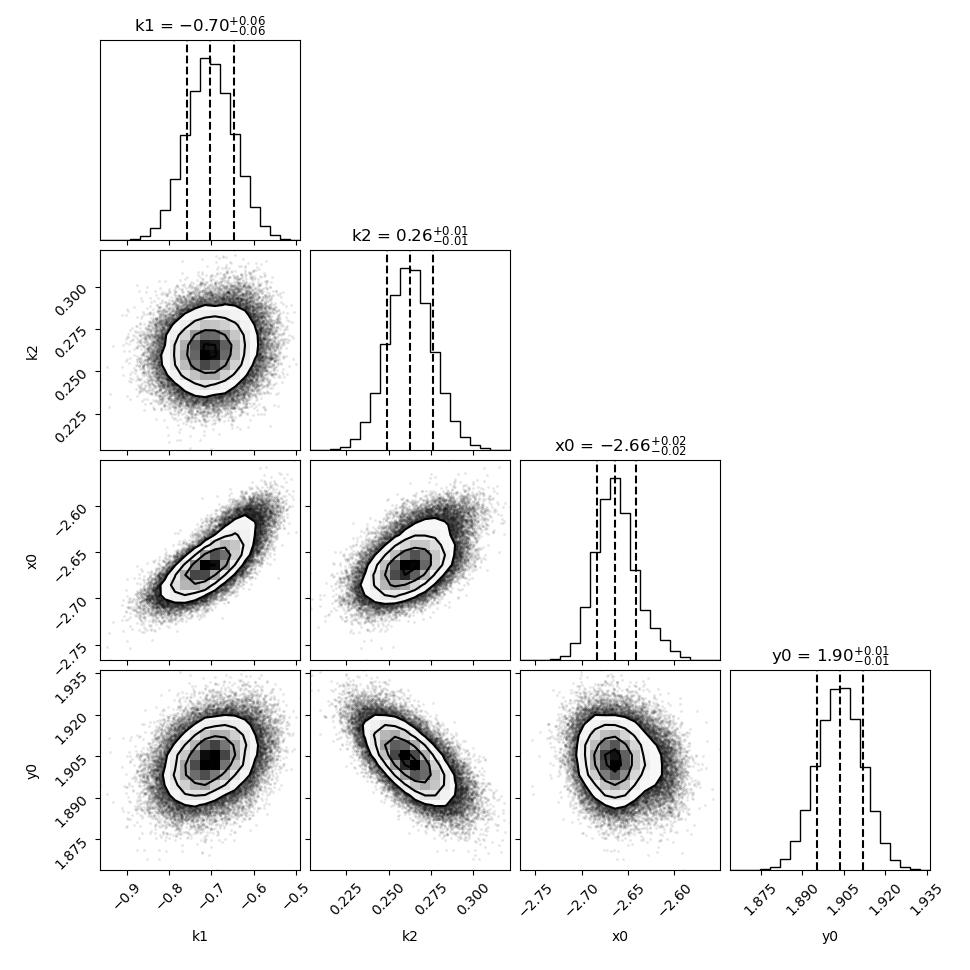}
    \end{subfigure}
   \\
    \makebox[\textwidth]{\footnotesize(2) H 1743-322 all}
    \caption{The results of fitting the multiple outbursts data of GX 339-4 and H 1743-322 (left figure) together and the corresponding corner plot (right figure).}
    \label{f6}
\end{figure*}

\begin{figure}[htbp]
\centering
\plotone{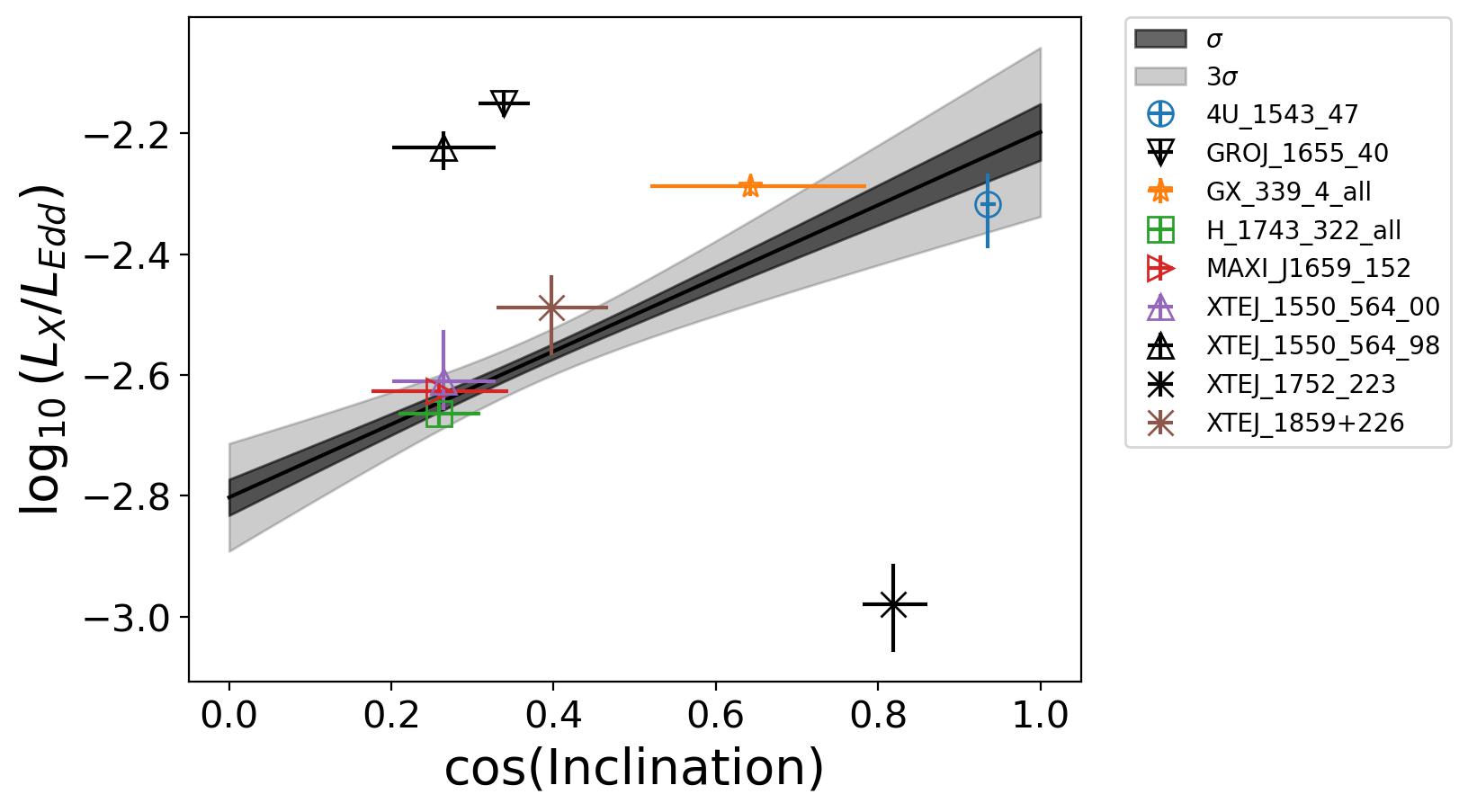}
\caption{The results obtained by fitting the multiple outbursts data of GX 339-4 and H 1743-322 together were used for the final linear fitting results. Different shapes of black dots represent the points that do not participate in the fitting. 
}
\label{f7}
\end{figure}

Figures \ref{f6} (1) and (2) respectively show the results of fitting the data of the four outbursts of GX 339-4 and H 1743-322 together and the corresponding corners. It can be seen that the increase in the overall number of data points improves the effect of curve fitting; moreover, the coordinates of the fitted cross points are almost the same as those obtained by separately fitting the $\Gamma - \xi_X$ evolution diagrams for each outburst. This also provides indirect support for the idea that for a single source, it is better to fit the data from multiple outbursts together for a more effective fitting result. Especially for GX 339-4, for some outbursts, their $\Gamma - \xi_X$ evolution diagram's have data at the cross point, which precisely fills in the gaps that occurred in the $\Gamma - \xi_X$ evolution diagram's cross point area due to the lack of data in some other outbursts. Figure \ref{f7} shows the linear relationship fitting results of replacing the results of separately calculating each outburst of GX 339-4 and H 1743-322 during linear fitting with the results of fitting all the outbursts of each source together. (The 2005 outburst of source GRO J1655-40, the 2009 outburst of source XTE J1752-223, and the 1998 outburst of source XTE J1550-564 do not participate in the linear relationship fitting.) The new linear relation between the inclination and $ \rm \xi_{X_{cross \ point}} $ is : $ \rm log_{10}(\xi_{X_{cross \ point}}) = 0.61^{+0.07}_{-0.07}\cos{\theta} - 2.80^{+0.03}_{-0.03} $, where $ \rm \theta $ is the inclination of BH XRB sources. We can see that the new linear fitting result is not much different from the initial linear fitting result and also supports the physical image we have mentioned in Section \ref{s4.1}.

\subsubsection{The 2005 outburst of source GRO J1655-40}

GRO J1655-40 is a well-known microquasar with an extremely high inclination ($\sim 70^{\circ}$) and a significantly relativistically broadened Fe line \citep{2001ApJ...554.1290G,DazTrigo2006}. \citet{2008BASI...36..151D} found that during the hard state of the outburst of this source in 2005, additional cooling Comptonization components were required to obtain a satisfactory spectral fit, especially in the early stage of the descending phase. This additional cooling mechanism may originate from a stronger soft photon field in the disk. We think that this potentially stronger soft photon field would directly enhance the X-ray flux we observe, causing the entire $\Gamma - \xi_X$ evolution diagram to shift overall to a higher $\xi_X$ value, ultimately resulting in a larger $ \rm \xi_{X_{cross \ point}} $ value in the fitted result. Additionally, during our data processing, we found that for the outburst of source GRO J1655-40 in 2005, the shape of its positive correlation branch is more inclined toward a cubic function rather than a simple linear function. This indicates that for the outburst of source GRO J1655-40 in 2005, the photon spectral index may have a more complex evolution with the X-ray Eddington ratio, which awaits further research in the future.

\subsubsection{The 1998 outburst of source XTE J1550-564}
\label{s4.4.3}

The 1998-1999 outburst of XTE J1550-564 was exceptionally bright, reaching a peak flux of approximately 6.8 Crab ($\rm 2-10 \ KeV$) in a giant flare, and exhibiting a double-peaked light curve \citep{Sobczak2000}. This outburst reached a significantly higher accretion rate and entered the extreme state (VHS) \citep{Kubota2004}. \citet{Titarchuk2010} discovered that the energy spectrum evolution of this source's 1998 outburst was fundamentally different from that of the homologous 2000 outburst: In the 1998 outburst, the cutoff energy experienced a complete ``down–minimum–up" pattern, while in the 2000 outburst, only the down and minimum stages were observed. This indicates that the rising phase of the 1998 outburst reached a much higher accretion rate, where the body motion Comptonization (i.e., Compton scattering in the inner region convergence flow) may have played a dominant role in photon energy enhancement, rather than simple thermal Comptonization \citep{Laurent2010}. Under the combined effect of body motion Comptonization, thermal Comptonization and very high accretion rate, we will observe more X-ray radiation than in ordinary BH XRB systems that only have thermal Comptonization and normal accretion rate. This systematically raises the $\xi_X$ of the source, ultimately causing the larger $ \rm \xi_{X_{cross \ point}} $ value in the fitted result.

\subsubsection{The 2009 outburst of source XTE J1752-223}

The 2009-2010 outburst of XTE J1752-223 was characterized by its intense and complex jet activity \citep{Yang2010,Russell2011,Debnath2021}. \citet{Debnath2021} indicated that during the low-luminosity hard state, the jet of this source was dense, while during the high-luminosity IMS, it transformed into a discrete or clumpy form. The X-ray energy spectrum of this source during the LHS was characterized by a very hard photon index ($\Gamma \sim 1.5$) and a high cutoff energy \citep{Stiele2011}. In this state, nonthermal jet emission made a negligible contribution to the X-ray flux, especially when the Eddington ratio was low (i.e., the anticorrelation branch), due to the jet emission having different spectral characteristics from the corona Comptonization, the jet usually produced a harder spectrum, and its contribution would cause the corresponding $\Gamma$ at low $\xi_X$ to increase, further leading to a greater anticorrelation slope of the $\Gamma - \xi_X$ evolution diagram at the low $\xi_X$, while the positive correlation slope of the $\Gamma - \xi_X$ evolution diagram at the high $\xi_X$ was basically unaffected. The different changes in the slopes of the two branches caused the final fitted $ \rm \xi_{X_{cross \ point}} $ to be smaller.

\subsection{Insights and precautions on the corona geometry}
\label{s4.5}

Our results have important implications for the geometric structure of the corona in BH XRBs. If the corona has a disk-like (``sandwich'') geometry rather than a spherical geometry, the positive correlation observed between $\xi_{X_{cross\ point}}$ and $\cos\theta$ can be naturally explained. In a spherical corona, the radiation is nearly isotropic, and the inclination of the accretion disk has only a weak effect on the observed coronal radiation. In contrast, in a disk-like corona, the projection effect is significant, and the observed spectral evolution should strongly depend on the viewing angle. This picture is directly supported by the MONK simulations. \citet{Fan2025} compared the spectra and polarization properties of three corona geometries (``sandwich,'' spherical, and pillar). They found that the ``sandwich'' and spherical models can produce similar spectra, but the inclination has a much stronger effect on the ``sandwich'' model. In this geometry, the path length of a photon through the corona increases with inclination as $1/\cos i$, so that the effective optical depth experienced by the escaping photons is larger at higher inclination. This results in a systematic decrease in $\Gamma$ and in the disk fraction. Our result that $\xi_{X_{cross\ point}}$ is positively correlated with $\cos\theta$ is consistent with the conclusion of the MONK simulations that high inclination suppresses the coronal radiation. Our findings therefore provide indirect but nontrivial evidence that at least some BH XRBs host a disk-like corona, consistent with recent polarization and spectral studies \citep{Zhang2025,Nosirov2026}. Furthermore, our results connect the positive correlation between $\Gamma$ and the Eddington ratio discovered by \citet{Cao2009} in AGN, the $\Gamma$-$\xi_X$ dual-branch model established by \citet{Qiao2013} for BH XRBs, and the geometric projection effect of inclination within a single physical framework. The core of this framework is the accretion structure of ``disk + corona + ADAF'' and its apparent differences under different viewing angles.

However, we cautiously point out that the inclination is not necessarily the only factor determining the position of the cross point. Factors such as the overly strong soft photon field, body motion Comptonization, and non-thermal jet may have a significant influence on the position of the cross point. Through the errors of the best-fit lines in Figures \ref{f3} and \ref{f7}, we can also infer that other parameters such as magnetic field strength, disk truncation radius, mass accretion rate, and the nature of the companion star can also influence the position of the cross point to a certain extent through their impact on the corona physics and spectral evolution. Additionally, our explanation strongly relies on the assumption that the underlying physical processes controlling the $\Gamma - \xi_X$ relationship in all sources are the same; if different BH XRBs have fundamentally different corona configurations (for example, some have pillar corona, while others have extended plate-like corona \citep{Lucchini2023,Zhang2024,Fan2025,Nekrasov2025}, then these sources may have completely different photon spectral indices and X-ray Eddington luminosity evolution and the position of the cross point. This might also be the reason why the $ \rm \xi_{X_{cross \ point}} $ of those three bursts in Section \ref{s4.5} deviated from the best-fitting line.

\section{Conclusion}
We referred to the low-quality X-ray binary sample compiled by K. Alabarta et al(\citeyear{2021MNRAS.507.5507A}), and through energy spectrum fitting of these samples, we obtained parameters such as X-ray luminosity and $ \Gamma $. Then, through the $\Gamma - \xi_X $ evolution diagram fitting and linear relationship fitting, the following conclusions were obtained:
\begin{enumerate}
\item For multiple outbursts of a BH XRB, the values of the $ \xi_X $ at the cross points of the $\Gamma - \xi_X$ evolution diagrams during each descending phase are approximately the same.
\item For the outbursts of different BH XRBs, there is a significant positive correlation between the values of the $ \rm \xi_{X_{cross \ point}} $ of the $\Gamma - \xi_X$ evolution diagrams during the descending phase of the outbursts and the cosine value of the inclination of the source ($ \rm log_{10}(\xi_{X_{cross \ point}}) = 0.72^{+0.09}_{-0.09}\cos{\theta} - 2.81^{+0.03}_{-0.03} $, where $ \rm \theta $ is the inclination of BH XRB sources). If we combine the data of the X-ray Eddington ratios of multiple outbursts of the BH XRBs with the data of the evolution of the photon spectral index and perform a linear fit on the obtained  X-ray Eddington ratios of the cross point, the expression of the linear relationship will slightly change to: $ \rm log_{10}(\xi_{X_{cross \ point}}) = 0.61^{+0.07}_{-0.07}\cos{\theta} - 2.80^{+0.03}_{-0.03} $, where $ \rm \theta $ is the inclination of BH XRB sources.
\end{enumerate}

\noindent Therefore, we think that the geometric effect of the inclination has a significant impact on the evolution of the energy spectrum of the BH XRBs. And the inclination angle might be the key factor causing the variation of $ \rm \xi_{X_{cross \ point}} $, which ranges from $ 0.1\% $ to $ 10\% $, in different BH XRBs.

In subsequent related work, on the one hand, the established linear relationship between $ \rm \xi_{X_{cross \ point}} $ and $\cos \theta$ can be tested during the new BH XRB outburst. Observations with high-sensitivity X-ray telescopes (such as IXPE, NICER, and XRISM) can provide higher-quality energy spectra and better-constrained inclinations, thereby enabling more precise verification of this correlation; on the other hand, after determining that there is such a positive linear relationship between the cosine of the inclination of the BH XRB and the cross point of its $\Gamma - \xi_X$ evolution diagram, we can then use this linear relationship to impose certain restrictions on the inclination of some known BH masses and related energy spectra, but with poorly defined inclinations of BH XRB sources. 

\begin{acknowledgments}
This work is supported by the NSFC (12363005), the National SKA Program of China (2022SKA0130104), the Foundation of Guizhou Provincial Education Department ((2020)0030), the Scientific Research Project of the Guizhou Provincial Education (KY[2022]132, KY[2022]137), Major Science and Technology Program of Xinjiang Uygur Autonomous Region (2022A03013-4) and Projects of the Grassroots Science Popularization Action Plan of Guizhou Provincial Association for Science and Technology.
\end{acknowledgments}

\begin{contribution}
A.D. provided the initial ideas and concepts. W.H. carried out the energy spectrum fitting. S.J. conducted the $\Gamma - \xi_X$ evolution didgram and linear fitting, as well as the preliminary writing of the article. The final article was completed through the joint discussion among A.D., W.H. and S.J..
\end{contribution}

\facilities{\emph{RXTE} \citep{2006AdSpR..38.2959S}.}

\software{Xspec \citep{1996ASPC..101...17A}, SciPy \citep{2020SciPy-NMeth}, emcee \citep{emcee}}

\bibliography{sample702}{}

@article{Wu2008,
  title = {The X‐Ray Spectral Evolution in X‐Ray Binaries and Its Application to Constrain the Black Hole Mass of Ultraluminous X‐Ray Sources},
  volume = {682},
  ISSN = {1538-4357},
  url = {http://dx.doi.org/10.1086/588187},
  DOI = {10.1086/588187},
  number = {1},
  journal = {The Astrophysical Journal},
  publisher = {American Astronomical Society},
  author = {Wu,  Qingwen and Gu,  Minfeng},
  year = {2008},
  month = jul,
  pages = {212–217}
}

@article{Fender2012,
  title = {Stellar-Mass Black Holes and Ultraluminous X-ray Sources},
  volume = {337},
  ISSN = {1095-9203},
  url = {http://dx.doi.org/10.1126/science.1221790},
  DOI = {10.1126/science.1221790},
  number = {6094},
  journal = {Science},
  publisher = {American Association for the Advancement of Science (AAAS)},
  author = {Fender,  Rob and Belloni,  Tomaso},
  year = {2012},
  month = aug,
  pages = {540–544}
}

@ARTICLE{1973A&A....24..337S,
       author = {{Shakura}, N.~I. and {Sunyaev}, R.~A.},
        title = "{Black holes in binary systems. Observational appearance.}",
      journal = {\aap},
         year = 1973,
        month = jan,
       volume = {24},
        pages = {337-355},
       adsurl = {https://ui.adsabs.harvard.edu/abs/1973A&A....24..337S}
}

@article{Remillard2006,
  title = {X-Ray Properties of Black-Hole Binaries},
  volume = {44},
  ISSN = {1545-4282},
  url = {http://dx.doi.org/10.1146/annurev.astro.44.051905.092532},
  DOI = {10.1146/annurev.astro.44.051905.092532},
  number = {1},
  journal = {Annual Review of Astronomy and Astrophysics},
  publisher = {Annual Reviews},
  author = {Remillard,  Ronald A. and McClintock,  Jeffrey E.},
  year = {2006},
  month = sep,
  pages = {49–92}
}

@ARTICLE{2013FrPhy...8..630Z,
       author = {{Zhang}, Shuang-Nan},
        title = "{Black hole binaries and microquasars}",
      journal = {Frontiers of Physics},
         year = 2013,
        month = dec,
       volume = {8},
       number = {6},
        pages = {630-660},
          doi = {10.1007/s11467-013-0306-z},
archivePrefix = {arXiv},
       eprint = {1302.5485},
 primaryClass = {astro-ph.HE},
       adsurl = {https://ui.adsabs.harvard.edu/abs/2013FrPhy...8..630Z}
}

@article{Done2007,
  title = {Modelling the behaviour of accretion flows in X-ray binaries: Everything you always wanted to know about accretion but were afraid to ask},
  volume = {15},
  ISSN = {1432-0754},
  url = {http://dx.doi.org/10.1007/s00159-007-0006-1},
  DOI = {10.1007/s00159-007-0006-1},
  number = {1},
  journal = {A \& ARv},
  publisher = {Springer Science and Business Media LLC},
  author = {Done,  Chris and Gierliński,  Marek and Kubota,  Aya},
  year = {2007},
  month = aug,
  pages = {1–66}
}

@article{Abramowicz1995,
  title = {Thermal equilibria of accretion disks},
  volume = {438},
  ISSN = {1538-4357},
  url = {http://dx.doi.org/10.1086/187709},
  DOI = {10.1086/187709},
  journal = {The Astrophysical Journal},
  publisher = {American Astronomical Society},
  author = {Abramowicz,  Marek A. and Chen,  Xingming and Kato,  Shoji and Lasota,  Jean-Pierre and Regev,  Oded},
  year = {1995},
  month = jan,
  pages = {L37}
}

@article{Narayan1995,
  title = {Advection-dominated Accretion: Underfed Black Holes and Neutron Stars},
  volume = {452},
  ISSN = {1538-4357},
  url = {http://dx.doi.org/10.1086/176343},
  DOI = {10.1086/176343},
  journal = {The Astrophysical Journal},
  publisher = {American Astronomical Society},
  author = {Narayan,  Ramesh and Yi,  Insu},
  year = {1995},
  month = oct,
  pages = {710}
}

@article{Liu2007,
  title = {The Existence of Inner Cool Disks in the Low/Hard State of Accreting Black Holes},
  volume = {671},
  ISSN = {1538-4357},
  url = {http://dx.doi.org/10.1086/522619},
  DOI = {10.1086/522619},
  number = {1},
  journal = {The Astrophysical Journal},
  publisher = {American Astronomical Society},
  author = {Liu,  B. F. and Taam,  Ronald E. and Meyer‐Hofmeister,  E. and Meyer,  F.},
  year = {2007},
  month = dec,
  pages = {695–705}
}

@article{Liu2010,
  title = {THE EFFECT OF CORONAL RADIATION ON A RESIDUAL INNER DISK IN THE LOW/HARD SPECTRAL STATE OF BLACK HOLE X-RAY BINARY SYSTEMS},
  volume = {726},
  ISSN = {1538-4357},
  url = {http://dx.doi.org/10.1088/0004-637X/726/1/10},
  DOI = {10.1088/0004-637x/726/1/10},
  number = {1},
  journal = {The Astrophysical Journal},
  publisher = {American Astronomical Society},
  author = {Liu,  B. F. and Done,  C. and Taam,  Ronald E.},
  year = {2010},
  month = dec,
  pages = {10}
}

@article{Yuan2014,
  title = {Hot Accretion Flows Around Black Holes},
  volume = {52},
  ISSN = {1545-4282},
  url = {http://dx.doi.org/10.1146/annurev-astro-082812-141003},
  DOI = {10.1146/annurev-astro-082812-141003},
  number = {1},
  journal = {Annual Review of Astronomy and Astrophysics},
  publisher = {Annual Reviews},
  author = {Yuan,  Feng and Narayan,  Ramesh},
  year = {2014},
  month = aug,
  pages = {529–588}
}

@article{Liu2019,
  title = {Evolution of the hard X-ray photon index in black-hole X-ray binaries: hints for accretion physics},
  volume = {487},
  ISSN = {1365-2966},
  url = {http://dx.doi.org/10.1093/mnras/stz1622},
  DOI = {10.1093/mnras/stz1622},
  number = {4},
  journal = {Monthly Notices of the Royal Astronomical Society},
  publisher = {Oxford University Press (OUP)},
  author = {Liu,  Hao and Dong,  AiJun and Weng,  ShanShan and Wu,  Qingwen},
  year = {2019},
  month = jul,
  pages = {5335–5345}
}

@article{Orosz1998,
  title = {Orbital Parameters for the Soft X‐Ray Transient 4U 1543−47: Evidence for a Black Hole},
  volume = {499},
  ISSN = {1538-4357},
  url = {http://dx.doi.org/10.1086/305620},
  DOI = {10.1086/305620},
  number = {1},
  journal = {The Astrophysical Journal},
  publisher = {American Astronomical Society},
  author = {Orosz,  Jerome A. and Jain,  Raj K. and Bailyn,  Charles D. and McClintock,  Jeffrey E. and Remillard,  Ronald A.},
  year = {1998},
  month = may,
  pages = {375–384}
}

@article{Torres2020,
  title = {Delimiting the black hole mass in the X-ray transient MAXI J1659-152 with Hα spectroscopy},
  volume = {501},
  ISSN = {1365-2966},
  url = {http://dx.doi.org/10.1093/mnras/staa3786},
  DOI = {10.1093/mnras/staa3786},
  number = {2},
  journal = {Monthly Notices of the Royal Astronomical Society},
  publisher = {Oxford University Press (OUP)},
  author = {Torres,  M A P and Jonker,  P G and Casares,  J and Miller-Jones,  J C A and Steeghs,  D},
  year = {2020},
  month = dec,
  pages = {2174–2181}
}

@article{Orosz2011,
  title = {AN IMPROVED DYNAMICAL MODEL FOR THE MICROQUASAR XTE J1550–564},
  volume = {730},
  ISSN = {1538-4357},
  url = {http://dx.doi.org/10.1088/0004-637X/730/2/75},
  DOI = {10.1088/0004-637x/730/2/75},
  number = {2},
  journal = {The Astrophysical Journal},
  publisher = {American Astronomical Society},
  author = {Orosz,  Jerome A. and Steiner,  James F. and McClintock,  Jeffrey E. and Torres,  Manuel A. P. and Remillard,  Ronald A. and Bailyn,  Charles D. and Miller,  Jon M.},
  year = {2011},
  month = mar,
  pages = {75}
}

@article{YanesRizo2022,
  title = {A refined dynamical mass for the black hole in the X-ray transient XTE J1859+226},
  volume = {517},
  ISSN = {1365-2966},
  url = {http://dx.doi.org/10.1093/mnras/stac2719},
  DOI = {10.1093/mnras/stac2719},
  number = {1},
  journal = {Monthly Notices of the Royal Astronomical Society},
  publisher = {Oxford University Press (OUP)},
  author = {Yanes-Rizo,  I V and Torres,  M A P and Casares,  J and Motta,  S E and Muñoz-Darias,  T and Rodríguez-Gil,  P and Armas Padilla,  M and Jiménez-Ibarra,  F and Jonker,  P G and Corral-Santana,  J M and Fender,  R},
  year = {2022},
  month = oct,
  pages = {1476–1482}
}

@article{Xu2015,
  title = {Anisotropic radiation from accretion disc coronae in active galactic nuclei},
  volume = {449},
  ISSN = {0035-8711},
  url = {http://dx.doi.org/10.1093/mnras/stv290},
  DOI = {10.1093/mnras/stv290},
  number = {1},
  journal = {Monthly Notices of the Royal Astronomical Society},
  publisher = {Oxford University Press (OUP)},
  author = {Xu,  Ya-Di},
  year = {2015},
  month = mar,
  pages = {191–198}
}

@article{Uttley2005,
  title = {Non-linear X-ray variability in X-ray binaries and active galaxies},
  volume = {359},
  ISSN = {1365-2966},
  url = {http://dx.doi.org/10.1111/j.1365-2966.2005.08886.x},
  DOI = {10.1111/j.1365-2966.2005.08886.x},
  number = {1},
  journal = {Monthly Notices of the Royal Astronomical Society},
  publisher = {Oxford University Press (OUP)},
  author = {Uttley,  P. and McHardy,  I. M. and Vaughan,  S.},
  year = {2005},
  month = may,
  pages = {345–362}
}

@ARTICLE{2003MNRAS.345.1057M,
       author = {{Merloni}, Andrea and {Heinz}, Sebastian and {di Matteo}, Tiziana},
        title = "{A Fundamental Plane of black hole activity}",
      journal = {\mnras},
         year = 2003,
        month = nov,
       volume = {345},
       number = {4},
        pages = {1057-1076},
          doi = {10.1046/j.1365-2966.2003.07017.x},
archivePrefix = {arXiv},
       eprint = {astro-ph/0305261},
 primaryClass = {astro-ph},
       adsurl = {https://ui.adsabs.harvard.edu/abs/2003MNRAS.345.1057M}
}

@article{McHardy2006,
  title = {Active galactic nuclei as scaled-up Galactic black holes},
  volume = {444},
  ISSN = {1476-4687},
  url = {http://dx.doi.org/10.1038/nature05389},
  DOI = {10.1038/nature05389},
  number = {7120},
  journal = {Nature},
  publisher = {Springer Science and Business Media LLC},
  author = {McHardy,  I. M. and Koerding,  E. and Knigge,  C. and Uttley,  P. and Fender,  R. P.},
  year = {2006},
  month = dec,
  pages = {730–732}
}

@inbook{McHardy2009,
  title = {LNP},
  ISBN = {9783540769378},
  ISSN = {1616-6361},
  url = {http://dx.doi.org/10.1007/978-3-540-76937-8_8},
  DOI = {10.1007/978-3-540-76937-8_8},
  booktitle = {The Jet Paradigm},
  publisher = {Springer Berlin Heidelberg},
  author = {McHardy,  I.},
  year = {2009},
  month = oct,
  pages = {203–232}
}

@article{Heil2015,
  title = {Inclination-dependent spectral and timing properties in transient black hole X-ray binaries},
  volume = {448},
  ISSN = {0035-8711},
  url = {http://dx.doi.org/10.1093/mnras/stv240},
  DOI = {10.1093/mnras/stv240},
  number = {4},
  journal = {Monthly Notices of the Royal Astronomical Society},
  publisher = {Oxford University Press (OUP)},
  author = {Heil,  L. M. and Uttley,  P. and Klein-Wolt,  M.},
  year = {2015},
  month = mar,
  pages = {3348–3353}
}

@article{MuozDarias2013,
  title = {Inclination and relativistic effects in the outburst evolution of black hole transients},
  volume = {432},
  ISSN = {1365-2966},
  url = {http://dx.doi.org/10.1093/mnras/stt546},
  DOI = {10.1093/mnras/stt546},
  number = {2},
  journal = {Monthly Notices of the Royal Astronomical Society},
  publisher = {Oxford University Press (OUP)},
  author = {Muñoz-Darias,  T. and Coriat,  M. and Plant,  D. S. and Ponti,  G. and Fender,  R. P. and Dunn,  R. J. H.},
  year = {2013},
  month = may,
  pages = {1330–1337}
}

@ARTICLE{2021MNRAS.507.5507A,
       author = {{Alabarta}, K. and {Altamirano}, D. and {M{\'e}ndez}, M. and {C{\'u}neo}, V.~A. and {Vincentelli}, F.~M. and {Castro-Segura}, N. and {Garc{\'\i}a}, F. and {Luff}, B. and {Veledina}, A.},
        title = "{Failed-transition outbursts in black hole low-mass X-ray binaries}",
      journal = {\mnras},
         year = 2021,
        month = nov,
       volume = {507},
       number = {4},
        pages = {5507-5522},
          doi = {10.1093/mnras/stab2241},
archivePrefix = {arXiv},
       eprint = {2107.10035},
 primaryClass = {astro-ph.HE},
       adsurl = {https://ui.adsabs.harvard.edu/abs/2021MNRAS.507.5507A}
}

@ARTICLE{2006AdSpR..38.2959S,
       author = {{Swank}, J.~H.},
        title = "{The Rossi X-ray timing explorer: Capabilities, achievements and aims}",
      journal = {Advances in Space Research},
         year = 2006,
        month = jan,
       volume = {38},
       number = {12},
        pages = {2959-2963},
          doi = {10.1016/j.asr.2006.07.009},
       adsurl = {https://ui.adsabs.harvard.edu/abs/2006AdSpR..38.2959S}
}

@article{Dong2020,
  title = {The spin measurement of the black hole in 4U 1543-47 constrained with the X-ray reflected emission},
  volume = {493},
  ISSN = {1365-2966},
  url = {http://dx.doi.org/10.1093/mnras/staa606},
  DOI = {10.1093/mnras/staa606},
  number = {3},
  journal = {Monthly Notices of the Royal Astronomical Society},
  publisher = {Oxford University Press (OUP)},
  author = {Dong,  Yanting and García,  Javier A and Steiner,  James F and Gou,  Lijun},
  year = {2020},
  month = mar,
  pages = {4409–4417}
}

@ARTICLE{1998ApJ...499..375O,
       author = {{Orosz}, Jerome A. and {Jain}, Raj K. and {Bailyn}, Charles D. and {McClintock}, Jeffrey E. and {Remillard}, Ronald A.},
        title = "{Orbital Parameters for the Soft X-Ray Transient 4U 1543-47: Evidence for a Black Hole}",
      journal = {\apj},
         year = 1998,
        month = may,
       volume = {499},
       number = {1},
        pages = {375-384},
          doi = {10.1086/305620},
archivePrefix = {arXiv},
       eprint = {astro-ph/9712018},
 primaryClass = {astro-ph},
       adsurl = {https://ui.adsabs.harvard.edu/abs/1998ApJ...499..375O}
}

@ARTICLE{2004MNRAS.354..355J,
       author = {{Jonker}, P.~G. and {Nelemans}, G.},
        title = "{The distances to Galactic low-mass X-ray binaries: consequences for black hole luminosities and kicks}",
      journal = {\mnras},
         year = 2004,
        month = oct,
       volume = {354},
       number = {2},
        pages = {355-366},
          doi = {10.1111/j.1365-2966.2004.08193.x},
archivePrefix = {arXiv},
       eprint = {astro-ph/0407168},
 primaryClass = {astro-ph},
       adsurl = {https://ui.adsabs.harvard.edu/abs/2004MNRAS.354..355J}
}

@INPROCEEDINGS{2003IAUS..212..365O,
       author = {{Orosz}, Jerome A.},
        title = "{Inventory of black hole binaries}",
    booktitle = {A Massive Star Odyssey: From Main Sequence to Supernova},
         year = 2003,
       editor = {{van der Hucht}, Karel and {Herrero}, Artemio and {Esteban}, C{\'e}sar},
       series = {IAUS},
       volume = {212},
        month = jan,
        pages = {365},
          doi = {10.48550/arXiv.astro-ph/0209041},
archivePrefix = {arXiv},
       eprint = {astro-ph/0209041},
 primaryClass = {astro-ph},
       adsurl = {https://ui.adsabs.harvard.edu/abs/2003IAUS..212..365O}
}

@article{Shaposhnikov2010,
  title = {DISCOVERY AND MONITORING OF A NEW BLACK HOLE CANDIDATE XTE J1752–223 WITHRXTE: RMS SPECTRUM EVOLUTION,  BLACK HOLE MASS,  AND THE SOURCE DISTANCE},
  volume = {723},
  ISSN = {1538-4357},
  url = {http://dx.doi.org/10.1088/0004-637X/723/2/1817},
  DOI = {10.1088/0004-637x/723/2/1817},
  number = {2},
  journal = {The Astrophysical Journal},
  publisher = {American Astronomical Society},
  author = {Shaposhnikov,  Nikolai and Markwardt,  Craig and Swank,  Jean and Krimm,  Hans},
  year = {2010},
  month = oct,
  pages = {1817–1824}
}

@article{Ratti2012,
  title = {The black hole candidate XTE J1752−223 towards and in quiescence: optical and simultaneous X-ray-radio observations: XTE J1752−223: outburst decay and quiescence},
  volume = {423},
  ISSN = {0035-8711},
  url = {http://dx.doi.org/10.1111/j.1365-2966.2012.21071.x},
  DOI = {10.1111/j.1365-2966.2012.21071.x},
  number = {3},
  journal = {Monthly Notices of the Royal Astronomical Society},
  publisher = {Oxford University Press (OUP)},
  author = {Ratti,  E. M. and Jonker,  P. G. and Miller-Jones,  J. C. A. and Torres,  M. A. P. and Homan,  J. and Markoff,  S. and Tomsick,  J. A. and Kaaret,  P. and Wijnands,  R. and Gallo,  E. and \"{O}zel,  F. and Steeghs,  D. T. H. and Fender,  R. P.},
  year = {2012},
  month = may,
  pages = {2656–2667}
}

@ARTICLE{2018ApJ...864...25G,
       author = {{Garc{\'\i}a}, Javier A. and {Steiner}, James F. and {Grinberg}, Victoria and {Dauser}, Thomas and {Connors}, Riley M.~T. and {McClintock}, Jeffrey E. and {Remillard}, Ronald A. and {Wilms}, J{\"o}rn and {Harrison}, Fiona A. and {Tomsick}, John A.},
        title = "{Reflection Spectroscopy of the Black Hole Binary XTE J1752-223 in Its Long-stable Hard State}",
      journal = {\apj},
         year = 2018,
        month = sep,
       volume = {864},
       number = {1},
          eid = {25},
        pages = {25},
          doi = {10.3847/1538-4357/aad231},
archivePrefix = {arXiv},
       eprint = {1807.01949},
 primaryClass = {astro-ph.HE},
       adsurl = {https://ui.adsabs.harvard.edu/abs/2018ApJ...864...25G}
}

@ARTICLE{2023AdSpR..71.3508D,
       author = {{Debnath}, Dipak and {Chatterjee}, Kaushik and {Nath}, Sujoy Kumar and {Chang}, Hsiang-Kuang and {Bhowmick}, Riya},
        title = "{Properties of 2017-18 'failed' outburst of GX 339-4}",
      journal = {Advances in Space Research},
         year = 2023,
        month = apr,
       volume = {71},
       number = {8},
        pages = {3508-3520},
          doi = {10.1016/j.asr.2022.12.011},
archivePrefix = {arXiv},
       eprint = {2111.13403},
 primaryClass = {astro-ph.HE},
       adsurl = {https://ui.adsabs.harvard.edu/abs/2023AdSpR..71.3508D}
}

@INPROCEEDINGS{2011IAUS..275..327C,
       author = {{Chen}, Tao},
        title = "{The determination of the GX 339-4's mass based on its 2010 outburst}",
    booktitle = {Jets at All Scales},
         year = 2011,
       editor = {{Romero}, Gustavo E. and {Sunyaev}, Rashid A. and {Belloni}, Tomaso},
       series = {IAUS},
       volume = {275},
        month = feb,
        pages = {327-328},
          doi = {10.1017/S1743921310016339},
       adsurl = {https://ui.adsabs.harvard.edu/abs/2011IAUS..275..327C}
}

@ARTICLE{2019MNRAS.488.1026Z,
       author = {{Zdziarski}, Andrzej A. and {Zi{\'o}{\l}kowski}, Janusz and {Miko{\l}ajewska}, Joanna},
        title = "{The X-ray binary GX 339-4/V821 Ara: the distance, inclination, evolutionary status, and mass transfer}",
      journal = {\mnras},
         year = 2019,
        month = sep,
       volume = {488},
       number = {1},
        pages = {1026-1034},
          doi = {10.1093/mnras/stz1787},
archivePrefix = {arXiv},
       eprint = {1904.07803},
 primaryClass = {astro-ph.SR},
       adsurl = {https://ui.adsabs.harvard.edu/abs/2019MNRAS.488.1026Z}
}

@article{Nandi2018,
  title = {Accretion flow dynamics during 1999 outburst of XTE J1859+226—modeling of broadband spectra and constraining the source mass},
  volume = {363},
  ISSN = {1572-946X},
  url = {http://dx.doi.org/10.1007/s10509-018-3314-1},
  DOI = {10.1007/s10509-018-3314-1},
  number = {5},
  journal = {Astrophysics and Space Science},
  publisher = {Springer Science and Business Media LLC},
  author = {Nandi,  Anuj and Mandal,  S. and Sreehari,  H. and Radhika,  D. and Das,  Santabrata and Chattopadhyay,  I. and Iyer,  N. and Agrawal,  V. K. and Aktar,  R.},
  year = {2018},
  month = apr 
}

@ARTICLE{2022MNRAS.517.1476Y,
       author = {{Yanes-Rizo}, I.~V. and {Torres}, M.~A.~P. and {Casares}, J. and {Motta}, S.~E. and {Mu{\~n}oz-Darias}, T. and {Rodr{\'\i}guez-Gil}, P. and {Armas Padilla}, M. and {Jim{\'e}nez-Ibarra}, F. and {Jonker}, P.~G. and {Corral-Santana}, J.~M. and {Fender}, R.},
        title = "{A refined dynamical mass for the black hole in the X-ray transient XTE J1859+226}",
      journal = {\mnras},
         year = 2022,
        month = nov,
       volume = {517},
       number = {1},
        pages = {1476-1482},
          doi = {10.1093/mnras/stac2719},
archivePrefix = {arXiv},
       eprint = {2209.10395},
 primaryClass = {astro-ph.HE},
       adsurl = {https://ui.adsabs.harvard.edu/abs/2022MNRAS.517.1476Y}
}

@ARTICLE{2009ApJ...699..453S,
       author = {{Shaposhnikov}, Nickolai and {Titarchuk}, Lev},
        title = "{Determination of Black Hole Masses in Galactic Black Hole Binaries Using Scaling of Spectral and Variability Characteristics}",
      journal = {\apj},
         year = 2009,
        month = jul,
       volume = {699},
       number = {1},
        pages = {453-468},
          doi = {10.1088/0004-637X/699/1/453},
archivePrefix = {arXiv},
       eprint = {0902.2852},
 primaryClass = {astro-ph.HE},
       adsurl = {https://ui.adsabs.harvard.edu/abs/2009ApJ...699..453S}
}

@ARTICLE{2016MNRAS.463..615K,
       author = {{Kalemci}, E. and {Begelman}, M.~C. and {Maccarone}, T.~J. and {Din{\c{c}}er}, T. and {Russell}, T.~D. and {Bailyn}, C. and {Tomsick}, J.~A.},
        title = "{Wind, jet, hybrid corona and hard X-ray flares: multiwavelength evolution of GRO J1655-40 during the 2005 outburst rise}",
      journal = {\mnras},
         year = 2016,
        month = nov,
       volume = {463},
       number = {1},
        pages = {615-627},
          doi = {10.1093/mnras/stw2002},
archivePrefix = {arXiv},
       eprint = {1609.00458},
 primaryClass = {astro-ph.HE},
       adsurl = {https://ui.adsabs.harvard.edu/abs/2016MNRAS.463..615K}
}

@ARTICLE{2003MNRAS.339.1031S,
       author = {{Shahbaz}, T.},
        title = "{Determining the spectroscopic mass ratio in interacting binaries: application to X-Ray Nova Sco 1994}",
      journal = {\mnras},
         year = 2003,
        month = mar,
       volume = {339},
       number = {4},
        pages = {1031-1040},
          doi = {10.1046/j.1365-8711.2003.06258.x},
archivePrefix = {arXiv},
       eprint = {astro-ph/0211266},
 primaryClass = {astro-ph},
       adsurl = {https://ui.adsabs.harvard.edu/abs/2003MNRAS.339.1031S}
}

@ARTICLE{1995Natur.375..464H,
       author = {{Hjellming}, R.~M. and {Rupen}, M.~P.},
        title = "{Episodic ejection of relativistic jets by the X-ray transient GRO J1655 - 40}",
      journal = {\nat},
         year = 1995,
        month = jun,
       volume = {375},
       number = {6531},
        pages = {464-468},
          doi = {10.1038/375464a0},
       adsurl = {https://ui.adsabs.harvard.edu/abs/1995Natur.375..464H}
}

@ARTICLE{2001ApJ...554.1290G,
       author = {{Greene}, Jenny and {Bailyn}, Charles D. and {Orosz}, Jerome A.},
        title = "{Optical and Infrared Photometry of the Microquasar GRO J1655-40 in Quiescence}",
      journal = {\apj},
         year = 2001,
        month = jun,
       volume = {554},
       number = {2},
        pages = {1290-1297},
          doi = {10.1086/321411},
archivePrefix = {arXiv},
       eprint = {astro-ph/0101337},
 primaryClass = {astro-ph},
       adsurl = {https://ui.adsabs.harvard.edu/abs/2001ApJ...554.1290G}
}

@ARTICLE{2002ApJ...567.1057T,
       author = {{Titarchuk}, Lev and {Shrader}, C.~R.},
        title = "{Observational Signatures of Black Holes: Spectral and Temporal Features of XTE J1550-564}",
      journal = {\apj},
         year = 2002,
        month = mar,
       volume = {567},
       number = {2},
        pages = {1057-1066},
          doi = {10.1086/338661},
archivePrefix = {arXiv},
       eprint = {astro-ph/0111272},
 primaryClass = {astro-ph},
       adsurl = {https://ui.adsabs.harvard.edu/abs/2002ApJ...567.1057T}
}

@ARTICLE{2011ApJ...730...75O,
       author = {{Orosz}, Jerome A. and {Steiner}, James F. and {McClintock}, Jeffrey E. and {Torres}, Manuel A.~P. and {Remillard}, Ronald A. and {Bailyn}, Charles D. and {Miller}, Jon M.},
        title = "{An Improved Dynamical Model for the Microquasar XTE J1550-564}",
      journal = {\apj},
         year = 2011,
        month = apr,
       volume = {730},
       number = {2},
          eid = {75},
        pages = {75},
          doi = {10.1088/0004-637X/730/2/75},
archivePrefix = {arXiv},
       eprint = {1101.2499},
 primaryClass = {astro-ph.SR},
       adsurl = {https://ui.adsabs.harvard.edu/abs/2011ApJ...730...75O}
}

@ARTICLE{2011MNRAS.414..677C,
       author = {{Coriat}, M. and {Corbel}, S. and {Prat}, L. and {Miller-Jones}, J.~C.~A. and {Cseh}, D. and {Tzioumis}, A.~K. and {Brocksopp}, C. and {Rodriguez}, J. and {Fender}, R.~P. and {Sivakoff}, G.~R.},
        title = "{Radiatively efficient accreting black holes in the hard state: the case study of H1743-322}",
      journal = {\mnras},
         year = 2011,
        month = jun,
       volume = {414},
       number = {1},
        pages = {677-690},
          doi = {10.1111/j.1365-2966.2011.18433.x},
archivePrefix = {arXiv},
       eprint = {1101.5159},
 primaryClass = {astro-ph.HE},
       adsurl = {https://ui.adsabs.harvard.edu/abs/2011MNRAS.414..677C}
}

@ARTICLE{2017ApJ...834...88M,
       author = {{Molla}, Aslam Ali and {Chakrabarti}, Sandip K. and {Debnath}, Dipak and {Mondal}, Santanu},
        title = "{Estimation of Mass of Compact Object in H 1743-322 from 2010 and 2011 Outbursts using TCAF Solution and Spectral Index-QPO Frequency Correlation}",
      journal = {\apj},
         year = 2017,
        month = jan,
       volume = {834},
       number = {1},
          eid = {88},
        pages = {88},
          doi = {10.3847/1538-4357/834/1/88},
archivePrefix = {arXiv},
       eprint = {1611.01266},
 primaryClass = {astro-ph.HE},
       adsurl = {https://ui.adsabs.harvard.edu/abs/2017ApJ...834...88M}
}

@ARTICLE{2012ApJ...745L...7S,
       author = {{Steiner}, James F. and {McClintock}, Jeffrey E. and {Reid}, Mark J.},
        title = "{The Distance, Inclination, and Spin of the Black Hole Microquasar H1743-322}",
      journal = {\apjl},
         year = 2012,
        month = jan,
       volume = {745},
       number = {1},
          eid = {L7},
        pages = {L7},
          doi = {10.1088/2041-8205/745/1/L7},
archivePrefix = {arXiv},
       eprint = {1111.2388},
 primaryClass = {astro-ph.HE},
       adsurl = {https://ui.adsabs.harvard.edu/abs/2012ApJ...745L...7S}
}

@ARTICLE{2015ApJ...803...59D,
       author = {{Debnath}, Dipak and {Molla}, Aslam Ali and {Chakrabarti}, Sandip K. and {Mondal}, Santanu},
        title = "{Accretion Flow Dynamics of MAXI J1659-152 from the Spectral Evolution Study of its 2010 Outburst using the TCAF Solution}",
      journal = {\apj},
         year = 2015,
        month = apr,
       volume = {803},
       number = {2},
          eid = {59},
        pages = {59},
          doi = {10.1088/0004-637X/803/2/59},
archivePrefix = {arXiv},
       eprint = {1504.05312},
 primaryClass = {astro-ph.HE},
       adsurl = {https://ui.adsabs.harvard.edu/abs/2015ApJ...803...59D}
}

@article{Beloborodov1999,
  title = {Plasma Ejection from Magnetic Flares and the X-Ray Spectrum of Cygnus X-1},
  volume = {510},
  ISSN = {0004-637X},
  url = {http://dx.doi.org/10.1086/311810},
  DOI = {10.1086/311810},
  number = {2},
  journal = {\apjl},
  publisher = {American Astronomical Society},
  author = {Beloborodov,  Andrei M.},
  year = {1999},
  month = jan,
  pages = {L123–L126}
}

@INPROCEEDINGS{1996ASPC..101...17A,
       author = {{Arnaud}, K.~A.},
        title = "{XSPEC: The First Ten Years}",
    booktitle = {Astronomical Data Analysis Software and Systems V},
         year = 1996,
       editor = {{Jacoby}, George H. and {Barnes}, Jeannette},
       series = {ASPC},
       volume = {101},
        month = jan,
        pages = {17},
       adsurl = {https://ui.adsabs.harvard.edu/abs/1996ASPC..101...17A}
}

@ARTICLE{2020SciPy-NMeth,
  author  = {Virtanen, Pauli and Gommers, Ralf and Oliphant, Travis E. and
            Haberland, Matt and Reddy, Tyler and Cournapeau, David and
            Burovski, Evgeni and Peterson, Pearu and Weckesser, Warren and
            Bright, Jonathan and {van der Walt}, St{\'e}fan J. and
            Brett, Matthew and Wilson, Joshua and Millman, K. Jarrod and
            Mayorov, Nikolay and Nelson, Andrew R. J. and Jones, Eric and
            Kern, Robert and Larson, Eric and Carey, C J and
            Polat, {\.I}lhan and Feng, Yu and Moore, Eric W. and
            {VanderPlas}, Jake and Laxalde, Denis and Perktold, Josef and
            Cimrman, Robert and Henriksen, Ian and Quintero, E. A. and
            Harris, Charles R. and Archibald, Anne M. and
            Ribeiro, Ant{\^o}nio H. and Pedregosa, Fabian and
            {van Mulbregt}, Paul and {SciPy 1.0 Contributors}},
  title   = {{{SciPy} 1.0: Fundamental Algorithms for Scientific
            Computing in Python}},
  journal = {NaMet},
  year    = {2020},
  volume  = {17},
  pages   = {261--272},
  adsurl  = {https://rdcu.be/b08Wh},
  doi     = {10.1038/s41592-019-0686-2},
}

@article{emcee,
   author = {{Foreman-Mackey}, D. and {Hogg}, D.~W. and {Lang}, D. and {Goodman}, J.},
    title = {emcee: The MCMC Hammer},
  journal = {PASP},
     year = 2013,
   volume = 125,
    pages = {306-312},
   eprint = {1202.3665},
      doi = {10.1086/670067}
}

@ARTICLE{2022ApJ...933...55C,
       author = {{{\v{C}}emelji{\'c}}, Miljenko and {Yang}, Hai and {Yuan}, Feng and {Shang}, Hsien},
        title = "{Formation of Episodic Jets and Associated Flares from Black Hole Accretion Systems}",
      journal = {\apj},
         year = 2022,
        month = jul,
       volume = {933},
       number = {1},
          eid = {55},
        pages = {55},
          doi = {10.3847/1538-4357/ac70cc},
archivePrefix = {arXiv},
       eprint = {2205.10531},
 primaryClass = {astro-ph.HE},
       adsurl = {https://ui.adsabs.harvard.edu/abs/2022ApJ...933...55C}
}

@BOOK{1979rpa..book.....R,
    author = {{Rybicki}, George B. and {Lightman}, Alan P.},
    title = "{Radiative Processes in Astrophysics}",
    year = 1979,
    publisher = {Wiley},
    adsurl = {https://ui.adsabs.harvard.edu/abs/1979rpa..book.....R}
}

@article{Nttil2024,
  title = {Radiative plasma simulations of black hole accretion flow coronae in the hard and soft states},
  volume = {15},
  ISSN = {2041-1723},
  url = {http://dx.doi.org/10.1038/s41467-024-51257-1},
  DOI = {10.1038/s41467-024-51257-1},
  number = {1},
  journal = {Nature Communications},
  publisher = {Springer Science and Business Media LLC},
  author = {N\"{a}ttil\"{a},  Joonas},
  year = {2024},
  month = Aug 
}

@article{Qiao2013,
  title = {A MODEL FOR THE CORRELATION OF HARD X-RAY INDEX WITH EDDINGTON RATIO IN BLACK HOLE X-RAY BINARIES},
  volume = {764},
  ISSN = {1538-4357},
  url = {http://dx.doi.org/10.1088/0004-637X/764/1/2},
  DOI = {10.1088/0004-637x/764/1/2},
  number = {1},
  journal = {The Astrophysical Journal},
  publisher = {American Astronomical Society},
  author = {Qiao,  Erlin and Liu,  B. F.},
  year = {2013},
  month = Jan,
  pages = {2}
}

@article{Cao2009,
  title = {An accretion disc-corona model for X-ray spectra of active galactic nuclei},
  volume = {394},
  ISSN = {1365-2966},
  url = {http://dx.doi.org/10.1111/j.1365-2966.2008.14347.x},
  DOI = {10.1111/j.1365-2966.2008.14347.x},
  number = {1},
  journal = {Monthly Notices of the Royal Astronomical Society},
  publisher = {Oxford University Press (OUP)},
  author = {Cao,  Xinwu},
  year = {2009},
  month = Mar,
  pages = {207–213}
}

@article{Fan2025,
  title = {Possible Coronal Geometry in the Hard and Soft States of Black Hole X-Ray Binaries from MONK Simulations},
  volume = {995},
  ISSN = {1538-4357},
  url = {http://dx.doi.org/10.3847/1538-4357/ae1a6e},
  DOI = {10.3847/1538-4357/ae1a6e},
  number = {1},
  journal = {The Astrophysical Journal},
  publisher = {American Astronomical Society},
  author = {Fan,  Ningyue and Bambi,  Cosimo and Steiner,  James F. and Zhang,  Wenda},
  year = {2025},
  month = Dec,
  pages = {124}
}

@article{DazTrigo2006,
  title = {XMM-Newton and INTEGRAL spectroscopy of the microquasar GRO J1655–40 during its 2005 outburst},
  volume = {462},
  ISSN = {1432-0746},
  url = {http://dx.doi.org/10.1051/0004-6361:20065406},
  DOI = {10.1051/0004-6361:20065406},
  number = {2},
  journal = {A \& A},
  publisher = {EDP Sciences},
  author = {Díaz Trigo,  M. and Parmar,  A. N. and Miller,  J. and Kuulkers,  E. and Caballero-García,  M. D.},
  year = {2006},
  month = Nov,
  pages = {657–666}
}

@ARTICLE{2008BASI...36..151D,
       author = {{Debnath}, D. and {Chakrabarti}, Sandip K. and {Nandi}, A. and {Mandal}, S.},
        title = "{Spectral and timing evolution of GRO J1655-40 during its outburst of 2005}",
      journal = {Bulletin of the Astronomical Society of India},
         year = 2008,
        month = dec,
       volume = {36},
       number = {4},
        pages = {151-189},
          doi = {10.48550/arXiv.0902.3791},
archivePrefix = {arXiv},
       eprint = {0902.3791},
 primaryClass = {astro-ph.GA},
       adsurl = {https://ui.adsabs.harvard.edu/abs/2008BASI...36..151D}
}

@article{Sobczak2000,
  title = {Complete<i>RXTE</i>Spectral Observations of the Black Hole X‐Ray Nova XTE J1550−564},
  volume = {544},
  ISSN = {1538-4357},
  url = {http://dx.doi.org/10.1086/317229},
  DOI = {10.1086/317229},
  number = {2},
  journal = {The Astrophysical Journal},
  publisher = {American Astronomical Society},
  author = {Sobczak,  Gregory J. and McClintock,  Jeffrey E. and Remillard,  Ronald A. and Cui,  Wei and Levine,  Alan M. and Morgan,  Edward H. and Orosz,  Jerome A. and Bailyn,  Charles D.},
  year = {2000},
  month = Dec,
  pages = {993–1015}
}

@article{Kubota2004,
  title = {Understanding of X-Ray Spectra of Black Hole Binaries and Its Application to ULXs},
  volume = {155},
  ISSN = {0375-9687},
  url = {http://dx.doi.org/10.1143/PTPS.155.19},
  DOI = {10.1143/ptps.155.19},
  journal = {Progress of Theoretical Physics Supplement},
  publisher = {Oxford University Press (OUP)},
  author = {Kubota,  Aya and Makishima,  Kazuo and Done,  Chris},
  year = {2004},
  pages = {19–26}
}

@article{Titarchuk2010,
  title = {IMPLICATION OF THE OBSERVED SPECTRAL CUTOFF ENERGY EVOLUTION IN XTE J1550–564},
  volume = {724},
  ISSN = {1538-4357},
  url = {http://dx.doi.org/10.1088/0004-637X/724/2/1147},
  DOI = {10.1088/0004-637x/724/2/1147},
  number = {2},
  journal = {The Astrophysical Journal},
  publisher = {American Astronomical Society},
  author = {Titarchuk,  Lev and Shaposhnikov,  Nikolai},
  year = {2010},
  month = Nov,
  pages = {1147–1152}
}

@article{Laurent2010,
  title = {SPECTRAL INDEX AS A FUNCTION OF MASS ACCRETION RATE IN BLACK HOLE SOURCES: MONTE CARLO SIMULATIONS AND AN ANALYTICAL DESCRIPTION},
  volume = {727},
  ISSN = {1538-4357},
  url = {http://dx.doi.org/10.1088/0004-637X/727/1/34},
  DOI = {10.1088/0004-637x/727/1/34},
  number = {1},
  journal = {The Astrophysical Journal},
  publisher = {American Astronomical Society},
  author = {Laurent,  Philippe and Titarchuk,  Lev},
  year = {2010},
  month = Dec,
  pages = {34}
}

@article{Debnath2021,
  title = {Jet properties of XTE J1752−223 during its 2009–2010 outburst},
  volume = {504},
  ISSN = {1365-2966},
  url = {http://dx.doi.org/10.1093/mnras/stab1169},
  DOI = {10.1093/mnras/stab1169},
  number = {3},
  journal = {Monthly Notices of the Royal Astronomical Society},
  publisher = {Oxford University Press (OUP)},
  author = {Debnath,  Dipak and Chatterjee,  Kaushik and Chatterjee,  Debjit and Jana,  Arghajit and Chakrabarti,  Sandip K},
  year = {2021},
  month = Apr,
  pages = {4242–4251}
}

@inproceedings{Stiele2011,
  series = {Texas 2010},
  title = {X-ray spectral and timing investigations of XTE J1752-223},
  url = {http://dx.doi.org/10.22323/1.123.0032},
  DOI = {10.22323/1.123.0032},
  booktitle = {Proceedings of 25th Texas Symposium on Relativistic Astrophysics  — PoS(Texas 2010)},
  publisher = {Sissa Medialab},
  author = {Stiele,  Holger and Munoz-Darias,  Teo and Motta,  Sara and Belloni,  Tomaso},
  year = {2011},
  month = Jun,
  pages = {032},
  collection = {Texas 2010}
}

@article{Zhang2025,
  title = {Constraining the corona geometry of Cyg X-1 with broad-band spectrum and polarimetric analysis based on observations in 2022 May},
  volume = {78},
  ISSN = {2053-051X},
  url = {http://dx.doi.org/10.1093/pasj/psaf127},
  DOI = {10.1093/pasj/psaf127},
  number = {1},
  journal = {Publications of the Astronomical Society of Japan},
  publisher = {Oxford University Press (OUP)},
  author = {Zhang,  Sixuan and Mizuno,  Tsunefumi and Kawashima,  Tomohisa and Done,  Chris and Fukazawa,  Yasushi and Takahashi,  Hiromitsu and Komine,  Ryusei and Takebayashi,  Koudai and Ohsuga,  Ken},
  year = {2025},
  month = Dec,
  pages = {123–136}
}

@article{Nosirov2026,
  title = {Assessing Systematic Uncertainties from Spectral Reanalysis of Cyg X-1 with Different Coronal Geometries},
  volume = {1005},
  ISSN = {1538-4357},
  url = {http://dx.doi.org/10.3847/1538-4357/ae75ab},
  DOI = {10.3847/1538-4357/ae75ab},
  number = {1},
  journal = {The Astrophysical Journal},
  publisher = {American Astronomical Society},
  author = {Nosirov,  Abdurakhmon and Jiang,  Jiachen and Bambi,  Cosimo and Tomsick,  John A.},
  year = {2026},
  month = Jun,
  pages = {5}
}

@article{Nekrasov2025,
  title = {Relativistic reflection within an extended hot plasma geometry},
  volume = {704},
  ISSN = {1432-0746},
  url = {http://dx.doi.org/10.1051/0004-6361/202556012},
  DOI = {10.1051/0004-6361/202556012},
  journal = {A \& A},
  publisher = {EDP Sciences},
  author = {Nekrasov,  A. D. and Dauser,  T. and García,  J. A. and Walton,  D. J. and Fromm,  C. M. and Young,  A. J. and Baker,  F. J. E. and Joyce,  A. M. and K\"{o}nig,  O. and Licklederer,  S. and H\"{a}fner,  J. and Wilms,  J.},
  year = {2025},
  month = Dec,
  pages = {A129}
}

@article{Zhang2024,
  title = {Inferring the iron K emissivity profiles of accretion discs irradiated by extended coronae},
  volume = {532},
  ISSN = {1365-2966},
  url = {http://dx.doi.org/10.1093/mnras/stae1714},
  DOI = {10.1093/mnras/stae1714},
  number = {4},
  journal = {Monthly Notices of the Royal Astronomical Society},
  publisher = {Oxford University Press (OUP)},
  author = {Zhang,  Wenda and Dovčiak,  Michal and Bursa,  Michal and Svoboda,  Jiří and Karas,  Vladimír},
  year = {2024},
  month = Jul,
  pages = {3786–3796}
}

@article{Lucchini2023,
  title = {Investigating the Impact of Vertically Extended Coronae on X-Ray Reverberation Mapping},
  volume = {951},
  ISSN = {1538-4357},
  url = {http://dx.doi.org/10.3847/1538-4357/acd24f},
  DOI = {10.3847/1538-4357/acd24f},
  number = {1},
  journal = {The Astrophysical Journal},
  publisher = {American Astronomical Society},
  author = {Lucchini,  Matteo and Mastroserio,  Guglielmo and Wang,  Jingyi and Kara,  Erin and Ingram,  Adam and Garcia,  Javier and Dauser,  Thomas and van der Klis,  Michiel and K\"{o}nig,  Ole and Lewin,  Collin and Nathan,  Edward and Panagiotou,  Christos},
  year = {2023},
  month = Jun,
  pages = {19}
}

@article{Russell2011,
  title = {A late jet rebrightening revealed from multiwavelength monitoring of the black hole candidate XTE J1752−223★: A jet rebrightening in XTE J1752−223},
  volume = {419},
  ISSN = {0035-8711},
  url = {http://dx.doi.org/10.1111/j.1365-2966.2011.19839.x},
  DOI = {10.1111/j.1365-2966.2011.19839.x},
  number = {2},
  journal = {Monthly Notices of the Royal Astronomical Society},
  publisher = {Oxford University Press (OUP)},
  author = {Russell,  D. M. and Curran,  P. A. and Muñoz-Darias,  T. and Lewis,  F. and Motta,  S. and Stiele,  H. and Belloni,  T. and Miller-Jones,  J. C. A. and Jonker,  P. G. and O’Brien,  K. and Homan,  J. and Casella,  P. and Gandhi,  P. and Soleri,  P. and Markoff,  S. and Maitra,  D. and Gallo,  E. and Cadolle Bel,  M.},
  year = {2011},
  month = Nov,
  pages = {1740–1751}
}

@article{Yang2010,
  title = {A decelerating jet observed by the EVN and VLBA in the X-ray transient XTE J1752−223},
  volume = {409},
  ISSN = {1745-3925},
  url = {http://dx.doi.org/10.1111/j.1745-3933.2010.00948.x},
  DOI = {10.1111/j.1745-3933.2010.00948.x},
  number = {1},
  journal = {MNRAS},
  publisher = {Oxford University Press (OUP)},
  author = {Yang,  J. and Brocksopp,  C. and Corbel,  S. and Paragi,  Z. and Tzioumis,  T. and Fender,  R. P.},
  year = {2010},
  month = Nov,
  pages = {L64–L68}
}

@article{Vincentelli2026,
  title = {The strength of Type-C quasi-periodic oscillations in black hole X-ray binaries correlates with the jet inclination},
  volume = {549},
  ISSN = {1365-2966},
  url = {http://dx.doi.org/10.1093/mnras/stag1004},
  DOI = {10.1093/mnras/stag1004},
  number = {4},
  journal = {Monthly Notices of the Royal Astronomical Society},
  publisher = {Oxford University Press (OUP)},
  author = {Vincentelli,  F M and Bollemeijer,  N and Veledina,  A and Altamirano,  D and Bu,  Q and Carotenuto,  F and Casella,  P and Cavecchi,  Y and Ma,  R and Marcel,  G and Mastroserio,  G and Motta,  S and Zhang,  L},
  year = {2026},
  month = May 
}

@article{Schnittman2006,
  title = {A Precessing Ring Model for Low‐Frequency Quasi‐periodic Oscillations},
  volume = {642},
  ISSN = {1538-4357},
  url = {http://dx.doi.org/10.1086/500923},
  DOI = {10.1086/500923},
  number = {1},
  journal = {The Astrophysical Journal},
  publisher = {American Astronomical Society},
  author = {Schnittman,  Jeremy D. and Homan,  Jeroen and Miller,  Jon M.},
  year = {2006},
  month = May,
  pages = {420–426}
}

@article{Varniere2016,
  title = {Impact of inclination on quasi-periodic oscillations from spiral structures},
  volume = {591},
  ISSN = {1432-0746},
  url = {http://dx.doi.org/10.1051/0004-6361/201527711},
  DOI = {10.1051/0004-6361/201527711},
  journal = {A \& A},
  publisher = {EDP Sciences},
  author = {Varniere,  P. and Vincent,  F. H.},
  year = {2016},
  month = Jun,
  pages = {A36}
}

@ARTICLE{1980A&A....86..121S,
       author = {{Sunyaev}, R.~A. and {Titarchuk}, L.~G.},
        title = "{Comptonization of X-Rays in Plasma Clouds - Typical Radiation Spectra}",
      journal = {\aap},
         year = 1980,
        month = jun,
       volume = {86},
        pages = {121},
       adsurl = {https://ui.adsabs.harvard.edu/abs/1980A&A....86..121S}
}

@article{Marra2024,
  title = {IXPE observation confirms a high spin in the accreting black hole 4U 1957+115},
  volume = {684},
  ISSN = {1432-0746},
  url = {http://dx.doi.org/10.1051/0004-6361/202348277},
  DOI = {10.1051/0004-6361/202348277},
  journal = {A \& A},
  publisher = {EDP Sciences},
  author = {Marra,  L. and Brigitte,  M. and Rodriguez Cavero,  N. and Chun,  S. and Steiner,  J. F. and Dovčiak,  M. and Nowak,  M. and Bianchi,  S. and Capitanio,  F. and Ingram,  A. and Matt,  G. and Muleri,  F. and Podgorný,  J. and Poutanen,  J. and Svoboda,  J. and Taverna,  R. and Ursini,  F. and Veledina,  A. and De Rosa,  A. and García,  J. A. and Lutovinov,  A. A. and Mereminskiy,  I. A. and Farinelli,  R. and Gunji,  S. and Kaaret,  P. and Kallman,  T. and Krawczynski,  H. and Kan,  Y. and Hu,  K. and Marinucci,  A. and Mastroserio,  G. and Mikus̆incová,  R. and Parra,  M. and Petrucci,  P.-O. and Ratheesh,  A. and Soffitta,  P. and Tombesi,  F. and Zane,  S. and Agudo,  I. and Antonelli,  L. A. and Bachetti,  M. and Baldini,  L. and Baumgartner,  W. H. and Bellazzini,  R. and Bongiorno,  S. D. and Bonino,  R. and Brez,  A. and Bucciantini,  N. and Castellano,  S. and Cavazzuti,  E. and Chen,  C. and Ciprini,  S. and Costa,  E. and Del Monte,  E. and Di Gesu,  L. and Di Lalla,  N. and Di Marco,  A. and Donnarumma,  I. and Doroshenko,  V. and Ehlert,  S. R. and Enoto,  T. and Evangelista,  Y. and Fabiani,  S. and Ferrazzoli,  R. and Hayashida,  K. and Heyl,  J. and Iwakiri,  W. and Jorstad,  S. G. and Karas,  V. and Kislat,  F. and Kitaguchi,  T. and Kolodziejczak,  J. J. and La Monaca,  F. and Latronico,  L. and Liodakis,  I. and Maldera,  S. and Manfreda,  A. and Marin,  F. and Marscher,  A. P. and Marshall,  H. L. and Massaro,  F. and Mitsuishi,  I. and Mizuno,  T. and Negro,  M. and Ng,  C. Y. and O’Dell,  S. L. and Omodei,  N. and Oppedisano,  C. and Papitto,  A. and Pavlov,  G. G. and Peirson,  A. L. and Perri,  M. and Pesce-Rollins,  M. and Pilia,  M. and Possenti,  A. and Puccetti,  S. and Ramsey,  B. D. and Rankin,  J. and Roberts,  O. J. and Romani,  R. W. and Sgrò,  C. and Slane,  P. and Spandre,  G. and Swartz,  D. A. and Tamagawa,  T. and Tavecchio,  F. and Tawara,  Y. and Tennant,  A. F. and Thomas,  N. E. and Trois,  A. and Tsygankov,  S. S. and Turolla,  R. and Vink,  J. and Weisskopf,  M. C. and Wu,  K. and Xie,  F.},
  year = {2024},
  month = Apr,
  pages = {A95}
}

@article{Titarchuk2025,
  title = {Linear Polarization in a Black Hole,  4U 1957+115 and Its X-Ray Spectral Analysis},
  volume = {991},
  ISSN = {1538-4357},
  url = {http://dx.doi.org/10.3847/1538-4357/adf55a},
  DOI = {10.3847/1538-4357/adf55a},
  number = {1},
  journal = {The Astrophysical Journal},
  publisher = {American Astronomical Society},
  author = {Titarchuk,  Lev and Seifina,  Elena and Soffitta,  Paolo and Makhinya,  Diana},
  year = {2025},
  month = Sep,
  pages = {68}
}

@article{Liu2022,
  title = {Accretion around black holes: The geometry and spectra},
  volume = {25},
  ISSN = {2589-0042},
  url = {http://dx.doi.org/10.1016/j.isci.2021.103544},
  DOI = {10.1016/j.isci.2021.103544},
  number = {1},
  journal = {iScience},
  publisher = {Elsevier BV},
  author = {Liu,  B.F. and Qiao,  Erlin},
  year = {2022},
  month = Jan,
  pages = {103544}
}

@book{burnham2002model,
  title     = {Model Selection and Multimodel Inference: A Practical Information-Theoretic Approach},
  author    = {Burnham, Kenneth P. and Anderson, David R.},
  edition   = {2},
  year      = {2002},
  publisher = {Springer},
  address   = {New York},
  isbn      = {978-0-387-95364-9},
  doi       = {10.1007/b97636}
}

@article{Nandi2012,
  title = {Accretion flow dynamics during the evolution of timing and spectral properties of GX 339-4 during its 2010–11 outburst},
  volume = {542},
  ISSN = {1432-0746},
  url = {http://dx.doi.org/10.1051/0004-6361/201117844},
  DOI = {10.1051/0004-6361/201117844},
  journal = {A \& A},
  publisher = {EDP Sciences},
  author = {Nandi,  A. and Debnath,  D. and Mandal,  S. and Chakrabarti,  S. K.},
  year = {2012},
  month = Jun,
  pages = {A56}
}

@article{Bhowmick2021,
  title = {Relation between Quiescence and Outbursting Properties of GX 339-4},
  volume = {910},
  ISSN = {1538-4357},
  url = {http://dx.doi.org/10.3847/1538-4357/abe134},
  DOI = {10.3847/1538-4357/abe134},
  number = {2},
  journal = {The Astrophysical Journal},
  publisher = {American Astronomical Society},
  author = {Bhowmick,  Riya and Debnath,  Dipak and Chatterjee,  Kaushik and Nagarkoti,  Shreeram and Chakrabarti,  Sandip Kumar and Sarkar,  Ritabrata and Chatterjee,  Debjit and Jana,  Arghajit},
  year = {2021},
  month = Apr,
  pages = {138}
}

@article{Ruan2019,
  title = {The Analogous Structure of Accretion Flows in Supermassive and Stellar Mass Black Holes: New Insights from Faded Changing-look Quasars},
  volume = {883},
  ISSN = {1538-4357},
  url = {http://dx.doi.org/10.3847/1538-4357/ab3c1a},
  DOI = {10.3847/1538-4357/ab3c1a},
  number = {1},
  journal = {The Astrophysical Journal},
  publisher = {American Astronomical Society},
  author = {Ruan,  John J. and Anderson,  Scott F. and Eracleous,  Michael and Green,  Paul J. and Haggard,  Daryl and MacLeod,  Chelsea L. and Runnoe,  Jessie C. and Sobolewska,  Malgosia A.},
  year = {2019},
  month = Sep,
  pages = {76}
}

@article{Yang2014,
  title = {Correlation between the photon index and X-ray luminosity of black hole X-ray binaries and active galactic nuclei: observations and interpretation},
  volume = {447},
  ISSN = {0035-8711},
  url = {http://dx.doi.org/10.1093/mnras/stu2571},
  DOI = {10.1093/mnras/stu2571},
  number = {2},
  journal = {Monthly Notices of the Royal Astronomical Society},
  publisher = {Oxford University Press (OUP)},
  author = {Yang,  Qi-Xiang and Xie,  Fu-Guo and Yuan,  Feng and Zdziarski,  Andrzej A. and Gierliński,  Marek and Ho,  Luis C. and Yu,  Zhaolong},
  year = {2014},
  month = Dec,
  pages = {1692–1704}
}

@article{Zhang2019,
  title = {Constraining the Size of the Corona with Fully Relativistic Calculations of Spectra of Extended Coronae. I. The Monte Carlo Radiative Transfer Code},
  volume = {875},
  ISSN = {1538-4357},
  url = {http://dx.doi.org/10.3847/1538-4357/ab1261},
  DOI = {10.3847/1538-4357/ab1261},
  number = {2},
  journal = {The Astrophysical Journal},
  publisher = {American Astronomical Society},
  author = {Zhang,  Wenda and Dovčiak,  Michal and Bursa,  Michal},
  year = {2019},
  month = Apr,
  pages = {148}
}

@article{Narayan2005,
  title = {Inclination Effects and Beaming in Black Hole X‐Ray Binaries},
  volume = {623},
  ISSN = {1538-4357},
  url = {http://dx.doi.org/10.1086/428709},
  DOI = {10.1086/428709},
  number = {2},
  journal = {The Astrophysical Journal},
  publisher = {American Astronomical Society},
  author = {Narayan,  Ramesh and McClintock,  Jeffrey E.},
  year = {2005},
  month = Apr,
  pages = {1017–1025}
}

@article{Jahoda2006,
  title = {Calibration of the
                    <i>Rossi X‐Ray Timing Explorer</i>
                    Proportional Counter Array},
  volume = {163},
  ISSN = {1538-4365},
  url = {http://dx.doi.org/10.1086/500659},
  DOI = {10.1086/500659},
  number = {2},
  journal = {The Astrophysical Journal Supplement Series},
  publisher = {American Astronomical Society},
  author = {Jahoda,  Keith and Markwardt,  Craig B. and Radeva,  Yana and Rots,  Arnold H. and Stark,  Michael J. and Swank,  Jean H. and Strohmayer,  Tod E. and Zhang,  William},
  year = {2006},
  month = Apr,
  pages = {401–423}
}

@article{Ingram2019,
  title = {A review of quasi-periodic oscillations from black hole X-ray binaries: Observation and theory},
  volume = {85},
  ISSN = {1387-6473},
  url = {http://dx.doi.org/10.1016/j.newar.2020.101524},
  DOI = {10.1016/j.newar.2020.101524},
  journal = {New Astronomy Reviews},
  publisher = {Elsevier BV},
  author = {Ingram,  Adam R. and Motta,  Sara E.},
  year = {2019},
  month = Sep,
  pages = {101524}
}

@article{Fabian2000,
  title = {Broad Iron Lines in Active Galactic Nuclei},
  volume = {112},
  ISSN = {1538-3873},
  url = {http://dx.doi.org/10.1086/316610},
  DOI = {10.1086/316610},
  number = {775},
  journal = {Publications of the Astronomical Society of the Pacific},
  publisher = {IOP Publishing},
  author = {Fabian,  A. C. and Iwasawa,  K. and Reynolds,  C. S. and Young,  A. J.},
  year = {2000},
  month = Sep,
  pages = {1145–1161}
}

@ARTICLE{1995A&AS..109....9G,
       author = {{Gottwald}, M. and {Parmar}, A.~N. and {Reynolds}, A.~P. and {White}, N.~E. and {Peacock}, A. and {Taylor}, B.~G.},
        title = "{The EXOSAT GSPC iron line catalog.}",
      journal = {\aaps},
         year = 1995,
        month = jan,
       volume = {109},
        pages = {9-28},
       adsurl = {https://ui.adsabs.harvard.edu/abs/1995A&AS..109....9G}
}

@article{Akaike1974,
  title = {A new look at the statistical model identification},
  volume = {19},
  ISSN = {0018-9286},
  url = {http://dx.doi.org/10.1109/TAC.1974.1100705},
  DOI = {10.1109/tac.1974.1100705},
  number = {6},
  journal = {IEEE Transactions on Automatic Control},
  publisher = {Institute of Electrical and Electronics Engineers (IEEE)},
  author = {Akaike,  H.},
  year = {1974},
  month = Dec,
  pages = {716–723}
}
\bibliographystyle{aasjournalv7.1}

\end{document}